\documentclass[12pt]{article}
\usepackage{amsfonts}
\usepackage{eurosym}
\usepackage{graphicx, tikz, tikzscale}
\usepackage{booktabs}
\usepackage{float}
\usepackage{caption, subcaption}
\usepackage{amssymb}
\usepackage{comment}
\usepackage{amsmath, bbm, amsthm}
\usepackage{booktabs}
\usepackage{multirow}
\usepackage{enumerate}
\usepackage{color}
\usepackage[authoryear]{natbib}
\usepackage{hyperref}
\usepackage{setspace}

\newcommand\epigraph[2]{
\hfill{}\begin{minipage}{5.8in}{\begin{spacing}{0.9}
\noindent\textit{#1}\end{spacing}
\hfill{}{#2}}\vspace{1em}
\end{minipage}}

\hypersetup{colorlinks=false, linkcolor=red, citecolor=blue}

\begin{document}
\title{Systemic Risk in Financial Networks:  A Survey
}
\author{Matthew O. Jackson and Agathe Pernoud\thanks{%
Department of
Economics, Stanford University, Stanford, California 94305-6072 USA.
Jackson is also an
external faculty member of the Santa Fe Institute.
Email:  jacksonm@stanford.edu and agathep@stanford.edu.  We gratefully
acknowledge financial support under NSF grants SES-1629446 and SES-2018554.
We thank Agostino Capponi, Stephen Morris, and Carlos Ramirez for
helpful comments.
Forthcoming in the {\sl Annual Review of Economics}
}}
\date{December 2020
}
\maketitle

\begin{abstract}
We provide an overview of the relationship between financial networks and systemic risk.  We present a taxonomy of different types of systemic risk, differentiating between direct externalities between
financial organizations (e.g., defaults, correlated portfolios and firesales), and perceptions and feedback effects (e.g., bank runs, credit freezes).
We also discuss optimal regulation and bailouts, measurements of systemic risk and financial centrality, choices by banks'
regarding their portfolios and partnerships, and the changing nature of financial networks.


\end{abstract}

\setcounter{page}{0}\thispagestyle{empty}
\newpage

\epigraph{``The difficult task before market participants, policymakers, and regulators with systemic
risk responsibilities such as the Federal Reserve is to find ways to preserve the benefits of
interconnectedness in financial markets while managing the potentially harmful side effects.''}{ Janet Yellen  \citeyearpar{yellen2013interconnectedness} }

\section{Introduction}

International finance has grown dramatically in past decades, paralleling the growth in international trade.
For instance, the amount of investment around the world coming from foreign sources went from
26 trillion dollars in 2000 to over 132 trillion dollars in 2016, which represents more
than a third of the total level of world investments.\footnote{See \cite{lund2017global}.}
In addition, the financial sector is characterized by strong interdependencies -
so capital is not only circulating between countries, but also from one financial institution to another.
Using administrative data from the US Federal Reserve Bank, \cite*{duartej2017} estimate
that 23\% of the assets of bank holding companies come from within the US financial system,
as well as 48\% of their liabilities - {\sl almost half}.

Globalization, and its associated economies of scope and scale,
have paid enormous dividends in terms of increased peace and prosperity.
However, the associated increasingly interconnected financial network among ever-larger nodes
also paves the way for systemic risk.  Interdependencies between financial institutions can act as amplification
mechanisms, and create channels for a shock in one part of the system to spread widely, leading to
losses that are much larger than the initial changes in fundamentals.  These are not idle concerns,
as we witnessed in 2008 when exposure to a problematic mortgage market led to
key insolvencies in the US and elsewhere,
and to a broad financial crisis and prolonged recession.\footnote{For narratives of the crisis see
the US Congressional {\sl Financial Crisis Inquiry Report} of January 2011, as
well as \cite{glassermany2016} and \cite{jackson2019}.}

Financial markets are ripe with externalities as the fates of institutions depend upon each
other in a variety of ways.
At a most basic level, insolvencies involve substantial costs which are then passed on via defaults and drops in equity values, especially if left to cascade.
The externalities are clear: if one organization has poor judgment in its investments,
poorly managed business practices, or even just unusually bad luck, this ends up affecting the
values of its partners, and their partners; and in discontinuous ways.
There are also many other forms of externalities in financial networks including bank runs,  changes in asset values due to fire sales, inferences that investors make
about one institution based on the health of another, and credit freezes.\footnote{We do not directly address the issue of bubbles in this survey, but
one can find extensive treatment elsewhere (e.g., \cite{shiller2015}).}
Although some of these risks can be hedged, there are no markets for insurance against many of them.
The externalities mean that the system as a whole can experience crises that are much broader and costlier than the
independent failures that ignite them -- hence the term ``systemic risk.''

Many forms of systemic risk can be mitigated or even
avoided altogether via
appropriate oversight and judicious intervention.
However, this requires a detailed view of financial interdependencies, an understanding of their
consequences, as well as of the incentives that different parties in the network have.
These are the focus of what follows.

The growth in the study of networks over the past decades
 has provided us with tools to better understand systemic risk.\footnote{For detailed overviews of the
broader networks literature see \cite*{jackson2008,jackson2019}.  References on financial networks appear in an
early review by \cite*{summer2013},
and references to more recent papers appear throughout this survey.
Chapter 4 of \cite*{jackson2019} details a financial crisis and discusses
some key aspects of financial markets and policy prescriptions.}
It is an ideal time to provide a conceptual framework within which we
can organize the main insights.
In what follows, we draw a distinction
between two types of systemic risk: (i) contagion through various channels that generate externalities among
financial institutions (e.g., defaults, correlated portfolios,
and firesales), and (ii) self-fulfilling prophecies and feedback
effects (e.g., bank runs, credit freezes, equilibrium multiplicity).
We then discuss how each sort of risk depends on the
network of interdependencies.
Finally, we use this taxonomy to examine: how systemic risk is
affected by
banks' incentives to choose their investments and partners,
how to measure systemic risk and financial
centrality, and when and how to intervene or regulate.

Some background on empirical analyses and facts about financial networks appear in the 
Supplemental Online Appendix, along with an executive summary of this survey.


\section{A Taxonomy of Systemic Risk in Financial Networks}\label{types_risk}

Defaults and financial crises are as old as investment:
from the immense credit crunch under the emperor Tiberius in 33CE, to the repeated external defaults
by most countries involved in the Napoleonic wars,
and the recurring bank runs and panics of the nineteenth century.
The variety of ways that such crises erupt and play out (e.g., \cite{reinhartr2009}) calls for a taxonomy of
the externalities that lead to systemic problems.

We provide a two-layer taxonomy.   We first distinguish between (i) contagion through direct
externalities (e.g., when a default by one bank leads to distress for another, or a firesale of one bank's assets depresses the value of another bank)
and (ii) various feedback effects that allow for multiple equilibria and self-fulfilling prophecies (e.g., when beliefs about the poor condition of a bank become self-fulfilling as they lead investors to call in their loans).
Within these two types of systemic risk, there is a second layer of different ways
in which each can work.

Before presenting this taxonomy, we discuss what
constitutes a financial network under different scenarios.

\subsection{What constitutes a financial network?}

Financial networks are complex systems in which many institutions are interconnected
in various ways.

First and foremost, institutions are linked through financial contracts:
they lend to and borrow from each other to smooth idiosyncratic liquidity variations and meet deposit requirements; they collaborate on investment opportunities; and they operate in chains -- repackaging and reselling assets to each other. These networks of interdependencies are the focus of a large part of the literature (e.g., \cite{alleng2000, eisenbergn2001, elliottgj2014}).

Second, even when financial institutions are not transacting directly,
commonality in their exposures lead their values to be correlated.  This can be tracked via a network
in which a (weighted) link between two institutions captures the correlation between
their portfolios (\cite{acharya2007, allen2012asset, dieboldy2014, cabralesgv2017}).

There is also a burgeoning literature tying these different forms of interdependencies together.
In \cite{heipertz2019transmission}, banks trade outside and interbank assets, and prices adjust to clear the markets.
The network is the reduced-form relationships between banks' equity values in equilibrium:
the weighted edge from $i$ to $j$ captures the partial equilibrium effect of a drop in the value of $i$ on that of $j$, given induced shifts in trades and prices.

Although the nature of the financial network varies across models, they highlight the same fact: financial
interdependencies generate systemic risks.
A formal model of financial networks is thus useful to measure, predict, and trace the sources of systemic risk.
Hence, we introduce a framework that encompasses many in the literature, and allows us to distinguish between two types of systemic risk.

Let $N=\{1,\dots, n\}$ be a set of financial institutions.
We call them banks, but they should be understood more broadly as all the institutions in the financial system whose
actions affect others.

The network is characterized by a matrix $\mathbf{G}=(g_{ij})_{ij}$ such
that $g_{ij}$ captures the connection between bank $i$ and $j$.  A connection $g_{ij}$ can have multiple components, for instance each corresponding to a different type of financial contract.\footnote{Thus, it may be a multigraph;  or one can think of it as multiplexed - so layered networks.
The different types of contracts can interact, as shown by \cite{bardoscia2017} for UK bank data.}

A key object of interest is the vector of values associated with each institution, $V=(V_i)_i$,
accounting for all assets and liabilities, including any defaults and associated bankruptcy costs.
Because banks are interconnected, their values depend on each other.
The value of bank $i$ is a function of others' values, denoted by $V_i=F_i(V\mid \mathbf{G})$.
Banks' values are then the solution to a system of $n$ equations in $n$ unknowns, written as:
\begin{equation}\label{eq}V = F(V\mid \mathbf{G}).\end{equation}
Under some conditions -- in particular that $F( \cdot \mid \mathbf{G})$
is nondecreasing and bounded in $V$\footnote{That is, if $V'_i\geq V_i$ for all $i$,
then $F(V'|\mathbf{G})\geq F(V|\mathbf{G})$, and also that the set of feasible $V$'s is bounded above and
below.} -- Tarski's fixed point theorem applies, and there exists an ``equilibrium'':
values, consistent with each other, that solve (\ref{eq}).  So, the term ``equilibrium''
simply refers to coherent accounting, rather than to a fixed point of best responses or of some dynamic system.
 There can exist multiple equilibria, and the set of equilibria forms a complete lattice: there are maximum and minimum equilibria
 that take on the highest and lowest possible
  equilibrium values for all institutions simultaneously, which we call the ``best'' and ``worst''
  equilibria in what follows.\footnote{The best and worst equilibria can be found via a simple algorithm:
  Start with the maximum possible values $V^{\max}$ (or minimum to find the worst) and then iteratively apply the function $F$.  In many financial models,
  the convergence is fast (e.g., see \cite{eisenbergn2001,jacksonp2020})
  as the base asset values drive the equations, while with more arbitrary interdependencies, finding the equilibria can be much slower (e.g., see \cite{etessami2019tarski}).}

We distinguish between two sources of systemic risk that are generated by interdependencies between banks.
First, a change in the value of bank $i$ affects bank $j$ whose value changes by
$\partial F_j/\partial V_i$. This then affects the values of banks connected to $j$, and so on: a
change in one bank's value spreads through the network and has far reaching consequences.
This form of risk is the focus of much of the literature on financial contagion.
The second type of systemic risk stems from the multiplicity of equilibria and a possible shift from one equilibrium to another. Even in the absence of any
change in the values of fundamental investments, network interdependencies can lead to self-fulfilling
feedback effects whereby changes in beliefs become realized.
So the first type of systemic risk captures how a change in fundamentals can move through the
network -- formally, how much equilibrium values $V$ change in response to some initial
change in fundamentals, while keeping the equilibrium being considered
constant -- whereas the second type of systemic risk captures shifts between
equilibria.\footnote{The distinction between these two types of systemic risk is reminiscent
of the two views of financial crises brought forward in the literature: the business cycle
view and the panic view (\cite*{allen2007introduction}).
In the former crises are driven by changes in fundamentals, whereas the latter are
self-fulfilling prophecies that can be triggered solely via beliefs and behaviors.}
A contagion-based crisis is triggered by a change in fundamentals, whereas what
triggers an equilibrium shift can be more nebulous and is less well-understood.
In the context of financial networks, an equilibrium shift can be interpreted as a market
``freeze,'' which is likely to be driven by an increase in uncertainty that leads banks and others to
be less trusting of their counterparties.

In the rest of this section, we discuss these two different forms of systemic
risk, and identify corresponding externalities and market imperfections generating
inefficiencies.

\subsection{Contagion Through Network Interdependencies} \label{section_contagion}

We discuss direct transmission of distress via counterparty risk and commonality in exposures.

\paragraph{Cascades of Insolvencies.}
A canonical form of contagion is a cascade of insolvencies.   A bank gets low returns on its investments and cannot pay its
debts. As those liabilities are defaulted upon, this worsens the balance sheets of other institutions
leading some of them to become insolvent.   As more become insolvent, the values of others are further
depressed and this cascades through the network.

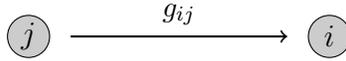
\begin{figure}[!h]
\begin{center}
\begin{tikzpicture}[scale=0.8]
\foreach \Point/\PointLabel in {(0,0)/j, (5,0)/i}
\draw[fill=black!20] \Point circle (0.35) node {$\PointLabel$};
\draw[->, thick] (0.7,0) to node [ above]  {\small $g_{ij}$} (4.3,0);
  \end{tikzpicture}
  \end{center}
  \captionsetup{singlelinecheck=off}
  \caption[]{\small Bank $j$ owes a debt of size $g_{ij}$ to $i$. 
  }\label{contagion}
\end{figure}

Consider the 2-bank relationship depicted in Figure \ref{contagion},
and let it represent a debt claim: $g_{ij}$ indicates that
 $j$ owes a debt $D_{ij}$ to $i$ - so arrows point in the direction in which value should flow.
In this example $g_{ij} = D_{ij}$. Interbank contracts generate interdependencies between
banks' (book) values:
\[V_i = F_i(V\mid \mathbf{G}) = \pi_i + \sum_j (d_{ij}(V) - D_{ji}), \]
where $\pi_i$ is the value of bank $i$'s portfolio of `outside'
investments,\footnote{These are investments
that are not in other financial institutions, such as mortgages, loans to non-financial companies, equity, etc. }
and $d_{ij}(V)$ the amount $j$ can manage to pay to $i$.
Because of limited liability, the value of this debt equals
\[d_{ij}(V) = \min\left\{D_{ij}, \frac{D_{ij}}{\sum_k D_{kj}}\left[\pi_j+\sum_k d_{jk}(V)\right]\right\}.\]
For simplicity we have equalized priority of all debt.
Here, $\pi_j+\sum_k d_{jk}(V)$ is the value that $j$ has available to
pay its debts, which is then divided across creditors in proportion to their claims on $j$; and more generally the function would be a nested function reflecting priorities.

To see how interdependencies generate systemic risk, let $D_{ij}=1$, $\pi_i=.5$ and suppose that $j$'s outside portfolio value drops from $\pi_j=1.5$ to $\pi_j'=.5$. Bank values start at $V_i=1.5, V_j=.5$ under $\pi$, but under $\pi'$ fall down to $V_i=1$ and $V_j=-.5$ (or effectively 0 given limited liability).
Though the shock only affected the portfolio of bank $j$, it
also depressed the value of the \emph{other} bank. The decrease in bank $i$'s value could then lead to
its default if $i$ had debts to others.

This sort of cascade does not lead to additional losses beyond the drop in portfolio value.
However, so far it ignores the fact that insolvencies involve substantial bankruptcy costs.
For instance, an extra $.5$ in bankruptcy costs would lead to $V_i=.5$ and $V_j=-1$ under $\pi'$.
Each additional insolvency then leads to deadweight losses to the system, and the overall cost can greatly exceed the initial shock.
Also lost are some of the investment and lending services of the insolvent banks.

Early models of counterparty risk include \cite*{rochett1996,alleng2000},
and model the behaviors of banks and depositors.
For example, \cite*{alleng2000} consider banks that are subject to
liquidity shocks (e.g., unanticipated withdrawls).
To insure against the shocks, banks can exchange part of their deposits ex ante. In the absence of
aggregate uncertainty, the first-best allocation can be implemented through cross-bank claims,
with a complete network of cross-deposits.  The banks that need early liquidity get it from banks
that have excess liquidity.  However, these claims generate financial instability and contagion upon
the realization of a shock that either was unanticipated, or  hits several banks, or when the
network is not appropriately connected.  Then, liquidity drawn by one bank from another can spillover and lead illiquidity to cascade.

\cite{eisenbergn2001} propose an algorithm to compute equilibrium payments between banks in a network of interbank debt liabilities.
The algorithm follows chains of defaults, and stops when no further default is induced by the previous ones.
More recent papers consider other types of financial contracts between banks, such
as equity claims (\cite*{elliottgj2014}). These claims make  market values of bank interdependent
as well: a drop in one's portfolio depresses its own value, which then depresses the value of its
equity holders, and their equity holders',
etc.\footnote{There is evidence that such equity-like cross-holdings generate
systemic risk. For instance, investment funds increasingly invest in each other,
and these cross-holdings have become a major source of vulnerability (\cite*{fricke2020connected}).}
Such models have been extended to include both debt and equity (\cite{jacksonp2019}).
Equity-like interdependencies have different implications for systemic risk than debt contracts,
as they enable banks to contribute to contagion without defaulting. For example,
Bank 1's drop in value, even if solvent,
can depress Bank 2's value if it holds shares of 1's stock, and
that can drive it to insolvency and incur bankruptcy costs.
This can precipitate defaults among Bank 2's creditors, especially if they were
already weak from holding Bank 1's stock.
Thus, combinations of drops in equity value, defaults on debt, and common exposures, can lead to cascades of defaults.

Inefficiency arises here from the externality that an institution's investment decision
affects the returns to others' portfolios and its ability to pay its debts in ways that cannot be
completely hedged by those affected.   These are not simply transfers of value from one institution to
another given that insolvency involves bankruptcy and other costs.

\paragraph{Correlated Investments, Fire Sales, and Other Exposures in Common.}
Another form of contagion, less direct, comes from externalities in asset prices. When a bank
becomes insolvent, it often has to sell, prematurely, significant amounts of assets in ``fire sales''.
Such dumping depresses prices for those assets, reducing the portfolio values of other banks
holding similar assets.
This can lead others to default, and their assets sales
to create a downward spiral
(\cite*{kiyotakim1997, cifuentesfs2005, gaik2010, greenwood2015vulnerable, capponi2015price}).
This is particularly problematic when portfolios are correlated across banks.
That leads both to stronger exposures, and greater pressures on prices in resulting fire sales.

The effect of fire sales on market prices depends on several market imperfections.   One is that the financial market is not deep enough to absorb a liquidation of a large bank's  portfolio
without a price impact.
There may also be asymmetric information, and market participants may infer something about underlying fundamentals when observing large-scale sales.
A decrease in market prices can thus amplify an initial shock, especially in a financial system in which
many assets are marked-to-market, there are asymmetries in information, and  large institutions.

Importantly, price-based contagion due commonalities in exposures across banks can
worsen cascades of insolvencies.
Here we see a three-level interaction between two counterparties who have similar exposures in their
investments.  First, they both tend to be vulnerable and near insolvency at the same time.
Second, if one is forced to sell off some of its assets, then the price effect can hurt the other's balance
sheet.
Third, if then one defaults on the other it can lead the second to become insolvent, especially in light of the first
two interactions which mean that it is already distressed.
Combined, these three effects can lead to cascades when either the direct default impact, common exposure, or indirect
price impact would not have led to further insolvencies by themselves.
For example, \cite{cifuentesfs2005} and \cite{gaik2010} show via simulations how contagion due to
counterparty risk can be amplified by fire sales. They consider financial networks that allow for
two types of linkages between banks: balance-sheet obligations, as well as price effects whenever
a bank is forced to de-leverage its portfolio. They then study how the risk of contagion depends
on the network structure of interbank obligations, and in particular its
density.\footnote{For the sake of simplicity, both papers assume that if one bank is forced to abruptly sell assets,
 the adverse effect on others' balance-sheets is the same for all the other banks.
How the interaction between these two networks affects the risk of contagion
under more general network structures remains a broadly open question.}
This is not just a theoretical concern, as there is evidence that two banks are much
more likely to be counterparties if their
 portfolios are more correlated (\cite{elliottgh2018}), suggesting that banks that are connected via
 financial obligations also tend to be more connected via commonalities in exposures.

\paragraph{Indirect Inferences.}
Commonalities in exposures pave the way for another form of contagion: ``guilt by similarity.''
People have doubts about the solvency of other enterprises that are similar to an insolvent one.\footnote{For instance, see \cite*{king1990transmission, acharya2008information, caballeros2013,alvarezb2015,stellian2020firms}.  For more general background on co-movement of firms' values as well as network positions, see \cite*{dieboldy2014}.}
Two key elements make such contagion possible: correlated portfolios across banks and uncertainty about the value of fundamentals and/or the banks' portfolio structures.

To illustrate these ideas, consider
$k=1,\dots K$ primitive assets, with independent values $p_k$.
A bank's portfolio is solely characterized by its investment in the different assets $q_i=(q_{ik})_k$.
To isolate the inference effect, suppose that banks have no contracts with each other: all of their value is based on their own portfolio of primitive assets.
The (undirected) financial network in this example captures correlation in asset holdings, such that $g_{ij} = \sum_k q_{ik}q_{jk} \sigma^2_k$, with $\sigma^2_k=\text{Var} (p_k)$.  Even if investors cannot directly observe the realized values of the different assets or the portfolios of the banks -- so they are unsure about either $q_i$s or the $p_k$s or both --  the market values of banks in equilibrium should still be consistent and satisfy
\[V_i = F_i(V\mid \mathbf{G}) = \mathbb{E}\left[ \sum_k q_{ik}p_k\mid V_{-i}\right] \quad\text{for all }i. \]
Consider once more the network in Figure \ref{contagion}, and let there be two outside assets with returns $p_A$ and $p_B$, respectively.
First, suppose that bank 2's entire portfolio is invested in asset $A$, that bank 1's is split equally between the two assets, and that this is known to investors.
Ex ante, without any additional information, the value of each bank simply equals the unconditional expected value of its portfolio: $V_1=\mathbb{E}[0.5(p_A+p_B)] = 0.5(\mu_A+\mu_B)$, and $V_2=\mu_A$.
Now consider what happens if it is revealed that Bank 1's value will be lower than expected: $0.5(p_A+p_B)=X<0.5(\mu_A+\mu_B)$.
Then, given the overlap in asset holdings, investors should update their valuation of Bank 2 as well to $V_2 = \mu_A -2\sigma_A^2(\sigma_A^2+\sigma_B^2)^{-1}[0.5(p_A+p_B)-X]<\mu_A$.

As a variation, suppose that $p_A=0$ and $p_B=1$ is known, and instead the
correlation comes from the fact that investors believe the two banks hold the same portfolio --
so they know that $q_1=q_2$, but not those values.   Then if they see that $V_1=.5$, they infer $V_2=.5$.

These correlations induce what we call ``inference-based contagion:'' upon observing a decrease in the value of Bank 1, investors make inferences about other banks' values
due to the correlation in portfolios (the source of the externality), both in terms of structure and payoffs, across banks.
With imperfect knowledge of those portfolios, they make inferences that could end up being justified ex post, or not. This form of inference-based contagion is made worse by the fact that banks are part of a complex financial network, whose structure is imperfectly known. \cite*{caballeros2013} show that the complexity of the network of interbank cross-exposures can lead risk-averse banks to take the prudential action more often than what is efficient, and to pull back funding from one another when a negative shock hits.
\bigbreak

These different types of externalities and interconnections interact and a
firm might be vulnerable in one network (e.g., inference from some other failure) and then cause a
cascade into another (e.g., then default on its payments).  Thus, proper evaluation of systemic risk
requires a holistic view of the different types of interdependencies between
institutions.\footnote{For more background on interconnected networks see \cite{kivelaetal2014,burkholz2016systemic,garas2016interconnected,atkisson2020understanding}.}

\subsection{Multiple Equilibria and Self-Fulfilling Feedback Effects}\label{selffeedback}

Systemic risk can arise even in the absence of any change in fundamental values. As soon as a financial
network allows for multiple equilibria, a mere shift in beliefs can move the system discontinuously
from one equilibrium to another, with real economic consequences.
Belief changes could arise from inferences, as mentioned above, that reflect real
underlying correlations; but they could also arise via sunspots (\cite{shell1989sunspot}), bubbles, or
exogenous events that can be conditioned upon by investors.  The key idea is that if there are multiple
equilibria, then which equilibrium applies depends on which one people expect.

\paragraph{Panics, and Runs.}
The classic form of bank runs and panics falls under this category of systemic risk, in which behavior
becomes self-fulfilling. This source of risk stems from  banks' primitive role of transforming short-term
deposits into long-term illiquid investments, which makes banks inherently fragile institutions: if
enough depositors withdraw their funding before the bank realizes its investments, the
bank cannot repay all of them and defaults. Classic treatments of this range from \cite*{keynes1936} to \cite*{diamondd1983},\footnote{For more background see \cite*{reinhartr2009}.} and show that by merely expecting a bank to be insolvent and withdrawing their deposits, depositors can induce its insolvency. 
Importantly, this sort of risk need not be triggered by a decrease in the value of the bank's fundamentals, but merely by a shift in beliefs about the health of the institution.  It could even be that people know that
a bank is healthy, but are worried that others are unsure of its health.%
\footnote{Of course, one can take this up to further levels of beliefs:  people might know the bank is healthy, and know that others know that the bank is healthy, but not know whether others know that everyone thinks the bank is healthy, etc. (\cite*{allen2006beauty,morris2002social}).}
The inefficiency here comes form the externality that returns on investment for a depositor depend on the behavior
of other depositors.
This complementarity in investments leads to the existence of
multiple equilibria, as people's expectations about how assets will be valued can come to be
self-fulfilling - and fear becomes contagious.\footnote{In some circumstances, one can refine the uncertainty and produce
unique predictions of self-fulfilling runs, as in \cite*{morriss1998}; but uncertainty about which
equilibrium will be played can also be problematic as in \cite*{rouknybs2018}. }
This holds even for an isolated bank, and hence even in the absence of any interdependencies between
financial institutions.\footnote{%
Non-depository institutions can also face similar liquidity risk: for instance,
broker-dealers can face runs from their collateral providers (\cite*{infante2018collateral}).}

\paragraph{Credit Freezes.}
Fear and pulling back of investments can occur not only on the part of depositors and outside investors, but also on the part of banks.
Uncertainty about economic conditions can lead banks to doubt how sound many businesses will be.  This can feed on itself, as if banks fear a recession they can pull back their capital and require ever higher interest rates.
This can lead to defaults, and banks to begin to doubt each other's health and to stop contracting with
each other, making it more difficult for banks to rebalance their portfolios.
This leads to further tightening and potential spiralling, and possibly to a complete credit freeze.
Again, this sort of freeze can be self-fulfilling:\footnote{The literature mostly highlights the
self-fulfilling, spiralling nature of credit freezes, which is why we mention it in this section,
but such behavior on the part of banks can also amplify the effect of a shock, and hence contribute to the first type of systemic risk as well. For instance, a bank hit by a  liquidity shortfall may need to withdraw  (or equivalently refuse to roll over) its loan to another bank to meet its payments, which may then be forced to call in its loans to others as well, etc.
An initially small liquidity shortfall can thus spread and lead to a broader liquidity crisis
(\cite{gai-haldane-kapadia}). }  the lack of investment worsens the conditions of businesses and
financial intermediaries, making them worse investments, which then justifies the pullback
(\cite{bebchuk2011self}); and thus this can be a problem even if no changes in fundamentals
drive the beliefs.
 This was present in the freeze of overnight lending between 2007 and 2009 (e.g., see the discussion in \cite*{brunnermeier2009,diamondr2011}.)
 Not only did lending dry up, but many stock markets around the world lost nearly half or more of their value (e.g, in the case of the Dow), while the underlying fundamentals did not reflect such a dramatic drop. Central banks had to provide much of the liquidity in interbank loan markets.

\begin{figure}[!h]
\begin{center}
\begin{tikzpicture}[scale=0.8]
\foreach \Point/\PointLabel in {(0,0)/1, (5,0)/2}
\draw[fill=black!20] \Point circle (0.35) node {$\PointLabel$};
\draw[->, thick] (0.7,0.3) [out=30,in=150]   to node [ above]  {\small $g_{21}$} (4.3,0.3);
\draw[<-, thick] (0.7,-0.3) [out=-30,in=-150]   to node [ below]  {\small $g_{12}$} (4.3,-0.3);
  \end{tikzpicture}
  \end{center}
  \captionsetup{singlelinecheck=off}
  \caption[]{\small A ring network with two banks. Arrows point in the direction in which value of claims flows.}\label{feedback}
\end{figure}
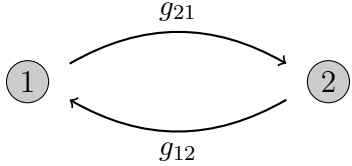

\paragraph{Self-Fulfilling Defaults.}
Financial contracts between banks can lead to self-fulfilling chains of defaults.
Recall that interbank contracts make bank values interdependent.
The anticipation of one bank failing to pay its debts can depress the value of other banks, and feedback to the original bank, making its default self-fulfilling.
Here we have externalities both in terms of payments and inferences.
As a simple example, consider the same model of interbank claims introduced in the section on cascades of insolvencies above, and the network depicted in Figure \ref{feedback}. For the sake of the example suppose neither bank has any outside assets, that they each owe $D_{12}=D_{21}=1$ to the other, and that the recovery rate on a defaulting bank's assets is zero. 
If one of the banks, say Bank 1, pays back its debt to the other, then Bank 2 has enough capital to pay its debt in full as well. Bank 1 is then indeed able to repay 2: such repayment is self-fulfilling, and there exists an equilibrium in which both banks remain solvent, and $V_1=V_2=0$. There exists however another equilibrium in which both banks default, and $V_1=V_2=-1$. Indeed, if Bank 1 expects not to have its claim on 2 paid back, then it cannot pay back its own debt, and vice versa.
Obviously, this example is trivial in the sense that the two banks should just cancel out each others' debts.  However, in more complicated cycles, especially ones with different forms of contracts and differing maturities,
such cancellation can be difficult to identify and execute.

This example shows how self-fulfilling default cascades
differ from classic bank runs as they are generated by network interdependencies,
rather than purely by beliefs.
They appear in any network of exposures between banks for which there are multiple equilibrium
values for interbank claims (e.g., \cite{elliottgj2014, rouknybs2018, jacksonp2020}).
When there are costs associated with bankruptcy, such cascades
are not just transfers failing to be made, but trigger real economic costs,
and this multiplicity of equilibria has
efficiency consequences.

\paragraph{Fire Sales and Contract Renegotiations.}
We close this section by highlighting that commonalities in asset holdings and fire sales can also generate multiple equilibria, and hence a self-fulfilling worsening of the financial system.
Consider, for instance, two banks holding the same asset, and suppose that the value of the asset drops
if a bank sells large quantities of it.
This could be due to either a lack of sufficient market depth, or from inferences if there is uncertainty
about why the asset is being sold.
 In normal times, if neither of the banks is forced to sell its holdings, the value of the asset remains high: there is an equilibrium with high bank values in which they both remain solvent. There is also another equilibrium in which they both dump a significant portion of their holdings, which depresses the price of the asset, and hence the values of the banks. This is self-fulfilling if one bank liquidating its holdings has a strong enough price impact to force the other to do so as well. This appears in the investment model of \cite{krishnamurthy2010amplification}, where multiple equilibria can coexist and exhibit various degrees of liquidation and price levels.  \cite{caballeros2013} consider a model that incorporates both dominos due to cross exposures between banks, and fire sales. They show that there can exist both an equilibrium in which contagion is contained and prices remain fair, and another in which banks take prudential actions, leading to fire sales, low market prices, and worse contagion.\footnote{See also  \cite{malherbe2014self}, who highlights the role of adverse selection in generating self-fulfilling liquidity dry-ups.}

Fire sales are not the only way in which the deterioration of banks'
balance sheets is exacerbated during stressed times.
So far we have taken obligations between banks as fixed, but they are
inherently dynamic and evolve. If a bank is expected to face low returns,
and as a result of being close to defaulting, then others will require
greater collateral when extending credit to it.
This worsens the situation of the bank, and can even precipitate its
default, making it self-fulfilling.%
\footnote{See \cite{fostel2008leverage,fostel2014endogenous}
for more background on the leverage cycle, which analyzes how much
collateral is required on loans and studies how it feeds back into asset prices in equilibrium.}

\bigskip

 Naturally, all these forms of systemic risk interact and are often at play at the same time.\footnote{\cite{siebenbrunner2020quantifying} discusses an approach to quantifying the relative
  contributions of different forces to systemic risk.}


\section{Network Structure and Systemic Risk}

We now discuss how network structure affects systemic risk.
Since, much of the literature on this question is based on networks of obligations between banks;
  we mostly restrict attention to
  interdependencies based on interbank contracts.

We first discuss how drops in equity values and defaults on debt can cascade, and how
such cascades depend on the structure of the network.
Second, we discuss how this type of contagion via counterparty risk is affected by additional interdependencies between banks stemming from correlated investments.
Third, we discuss systemic risk stemming from self-fulfilling feedbacks and on the kinds of network patterns that generate such feedbacks.

\subsection{Non-monotonicities in Network Density}\label{monotonicity}

Countervailing forces in financial networks lead contagion to be nonmonotonic in network density.    This is a point studied in detail by \cite*{elliottgj2014}, and it applies to a variety of models including \cite*{cifuentesfs2005, gaik2010, wagner2010, elliottgj2014, gofman2017,jacksonp2019}.
This distinguishes contagion in financial markets  from, for instance contagion of a disease or diffusion
of an idea, for which adding more interactions only leads to more extensive rates of spreading.%
\footnote{This also goes beyond what is known as `complex contagion':
contagion of something that takes several interactions to lead to infection --
for instance, hearing a rumor several times before believing it and passing it along,
or following the actions of a majority of friends.  See \cite{centola2018} and \cite{jackson2019}
for  background discussion and references,
and \cite{jacksons2017} for a detailed look at how complex contagion varies with network structure.
Financial networks have elements of complex contagion, since a bank may only become insolvent after
several counter-parties default, but also have nonmonotonicities in how many interactions they have. }

As a bank adds counterparties it becomes
susceptible to drops in values or defaults from more sources -
which tends to increase the potential for cascades.  However, holding a bank's total exposure
constant, spreading that exposure over more counterparties makes it less exposed to any given counterparty, which lowers the potential for contagion.
To study these two forces,    \cite*{elliottgj2014}
distinguish two basic dimensions of the interconnectivity between financial institutions:
how many partners each institution has, which they call the ``density'' of the network;
and the fraction of a bank's portfolio held in contracts with other institutions, which they call the ``integration'' of the network.

We illustrate the nonmonotonicity of these two forces in the context of a simple example.

\begin{figure}[!h]
\centering
\begin{subfigure}{.5\textwidth}
  \centering
  \includegraphics[width=1\linewidth]{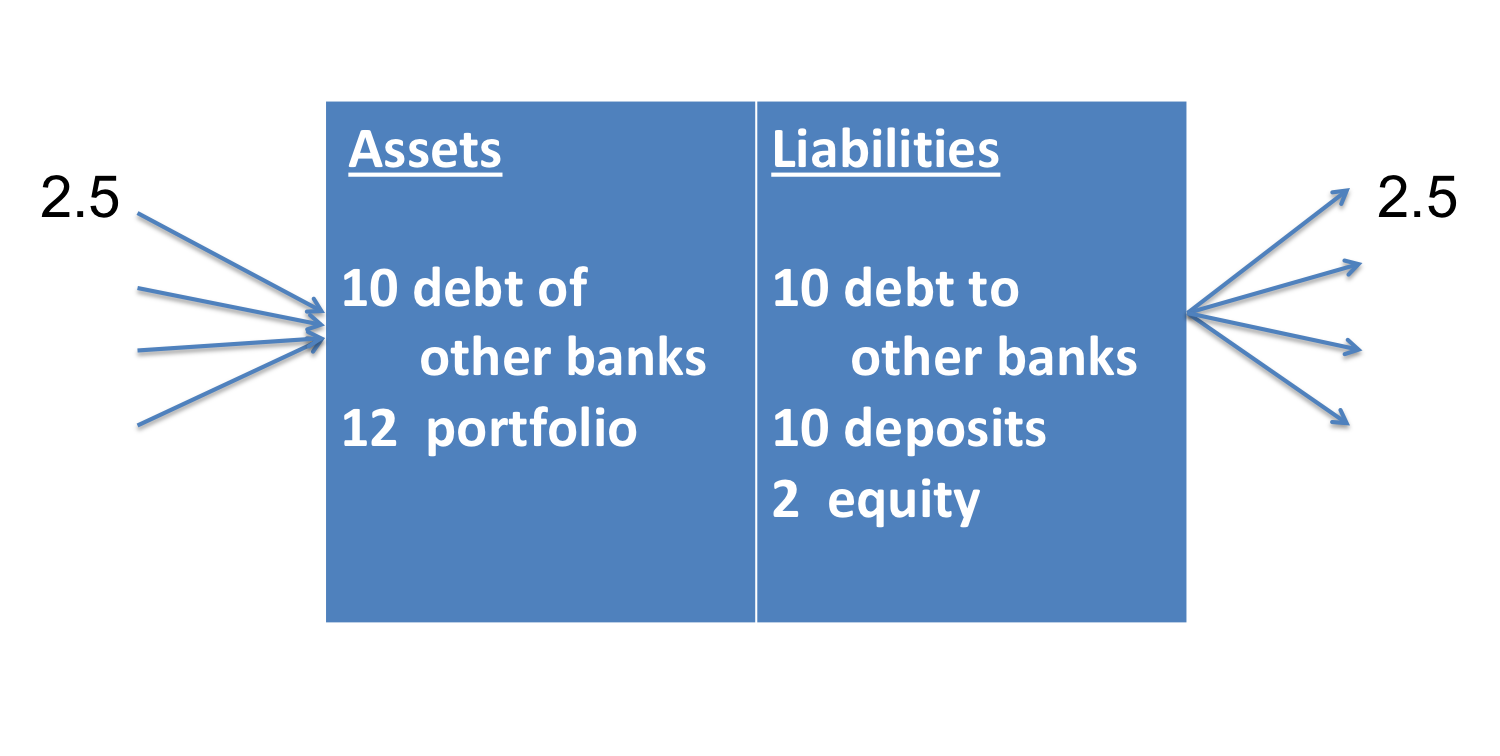}
\end{subfigure}%
\begin{subfigure}{.5\textwidth}
  \centering
  \includegraphics[width=1\linewidth]{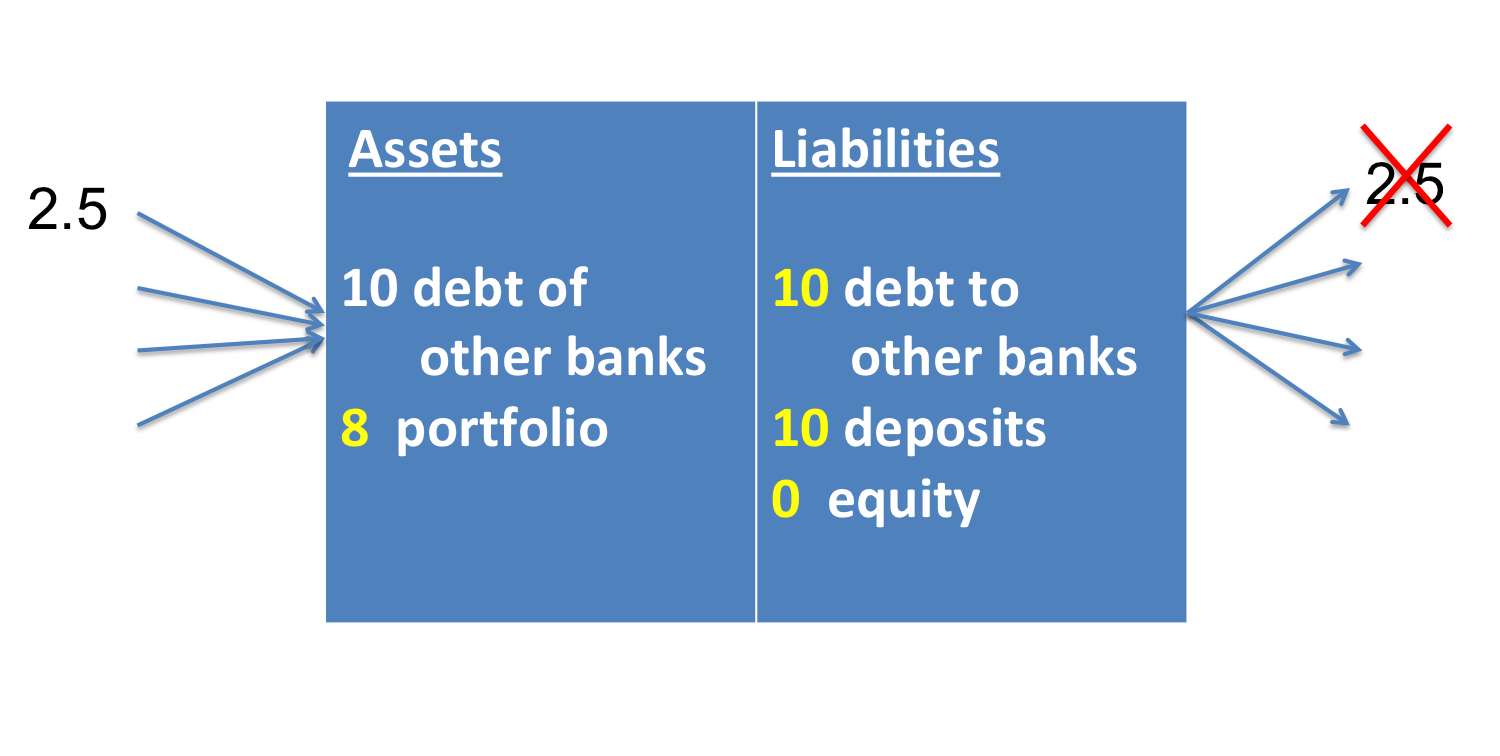}
\end{subfigure}
\caption{\small (Left) The starting balance sheet of the banks in a network. (Right) Some bank's
portfolio drops to a value below 10, say to 8.   This makes the bank insolvent, and so it
defaults on some of its payments.}\label{fig:nonmonotone1}
\end{figure}

Consider a network of identical banks that have balance sheets of the form in
Figure \ref{fig:nonmonotone1} (Left).

On the liability side, each bank has 10 units of capital from deposits, another 10 units of capital from loans from other banks, and their owners
have 2 units of capital in the form of equity.   On the asset side, each bank has
an investment portfolio worth 12 and loans to other banks worth 10.

In this example, we can measure the level of integration as 10 - which is how much of the bank's
assets comes from other banks, in this case in the form of interbank debt.
The density for this bank is 4, counting the number of counterparties the bank has.   So those levels of integration and density lead to
an exposure of 2.5 per counterparty.   In this example, the bank also owes 2.5 to each of four counterparties, so that there is a full symmetry.

Now let us suppose that the investment portfolio of one of these banks drops in value,
as in Figure \ref{fig:nonmonotone1} (Right).
This bank is now insolvent,
and so it defaults on some of its payments.
For the purposes of this example, let us treat the default as total on at least one of its loans, due to
bankruptcy costs,  although one can obviously extend the example to work with some partial default.

\begin{figure}[!h]
\centering
\begin{subfigure}{.5\textwidth}
  \centering
  \includegraphics[width=1\linewidth]{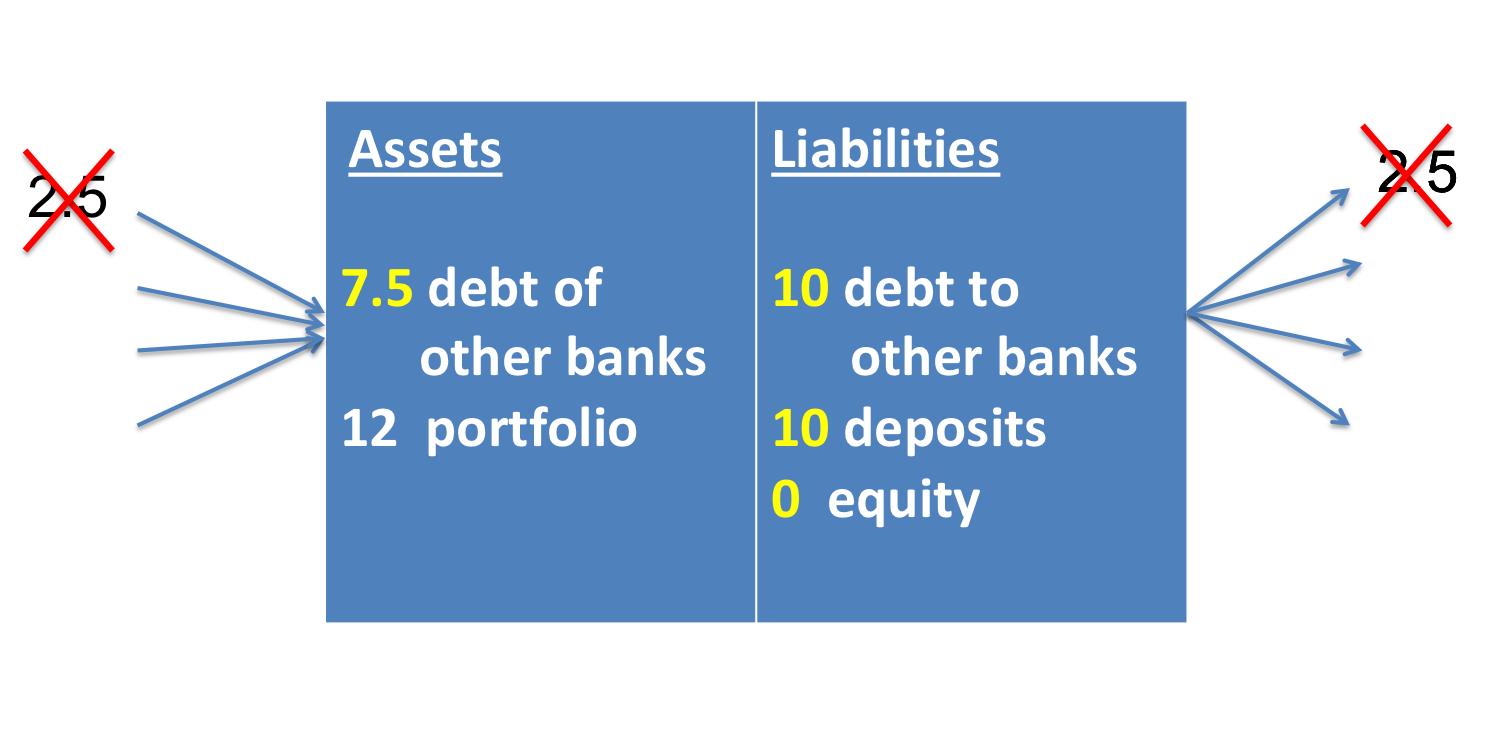}
\end{subfigure}%
\begin{subfigure}{.5\textwidth}
  \centering
  \includegraphics[width=1\linewidth]{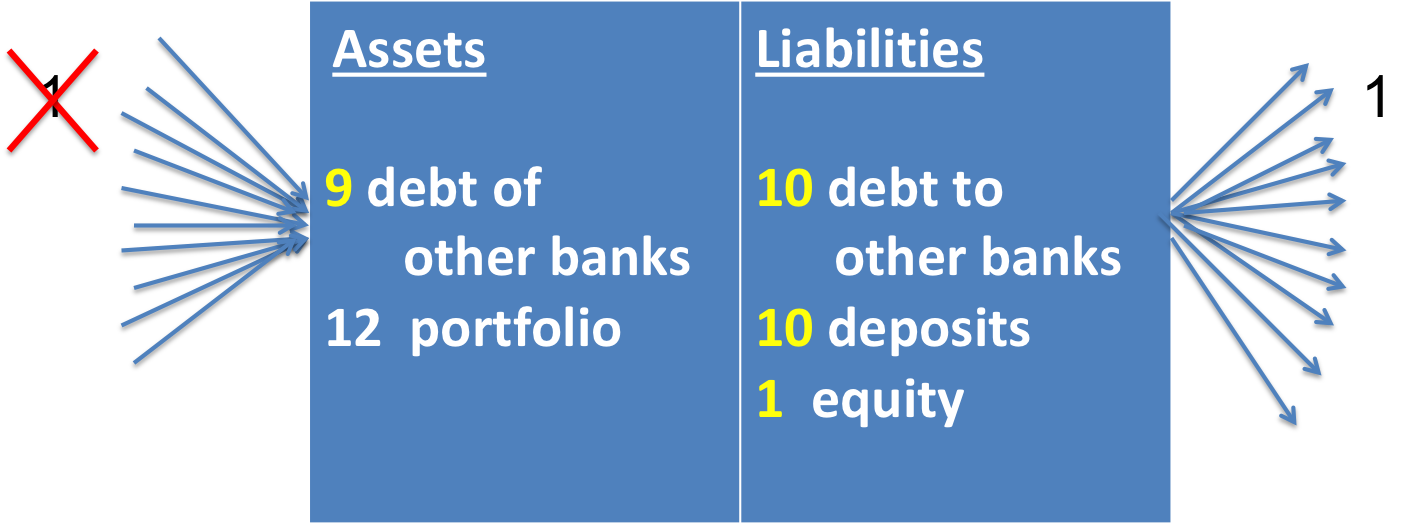}
\end{subfigure}
\caption{\small (Left) A second bank now becomes insolvent due to its lost asset value from the loan to the first bank.  It then defaults on some of its payments. (Right) Even though banks have more counterparties, the lower exposure to each separate one of them now makes them immune to a default by any single counterparty.}\label{fig:nonmonotone3}
\end{figure}

Initially the bank owed four different banks 2.5, and so it fails to make at least one of these payments.
This then has to be written off by the counterparty that made the loan to the first bank, and so that second bank loses 2.5 of its assets.
The second bank
is now insolvent as well and defaults on some of its payments.
Again, let us presume bankruptcy costs so that
it fails to pay any of at least one of its loans,  as pictured in Figure \ref{fig:nonmonotone3} (Left).
This now cascades.

With the exposure of 2.5 to each other bank, and an initial equity value of only 2, banks are susceptible
to even a single defaulting counterparty.
Both the level of integration and the density of the network in this example
are important in driving the defaults.
If the banks had lower levels of integration, but the same density, their exposure to any
given counterparty would be less than 2.5, and if it was less than 2, then no single counterparty's
default could erase a bank's equity value.
Increasing the integration -- i.e., the {\sl amount} of exposure of each bank to others -- tends to increase the
propensity for contagion, as it does in this example.\footnote{However,
as \cite*{elliottgj2014} also discuss, increasing integration can help diversify a given bank's
portfolio by changing the assets which it is implicitly holding through its connections --
depending on the circumstances.
Thus, more exposure can help diversify any given bank's portfolio, making
their own investments less variable and more stable, but also leads to
an increase in the possibility of contagion. }

Similarly, if a bank still had an integration level of 10, but
greater density (more partners) so that it had exposure of no more than 2 to any single counterparty,
then the cascade would also have been avoided.

To see the importance of density, let us alter the example so that each bank has 10 counterparties to
which it owes 1 unit each, as in Figure \ref{fig:nonmonotone3} (Right).
So, the level of integration is the same, but the density has increased.
In this case, there is no longer any cascade.    The default by any single counterparty no longer leads a bank to become insolvent.

Here we see the nonmonotonicity quite clearly.   We have increased the number of counterparties of each bank, and hence made the financial network denser, and yet have eliminated the cascade.     It is nonmonotonic since if we started with no counterparties, then there would not have been any contagion.
Or, if we just had two banks each paired to each other, then one would have dragged the other down, but it would not have spread.
The case with four counterparties hits a ``sweet spot'':   the density is high enough to lead to a very connected network where things can spread widely, and each bank is exposed enough to others that a single default can trigger an insolvency in a counterparty.
Once we increase up to ten counterparties, then a single default is no longer a major problem for any single counterparty.

\begin{figure}[!h]
\centering
{\includegraphics[width=0.6\textwidth]{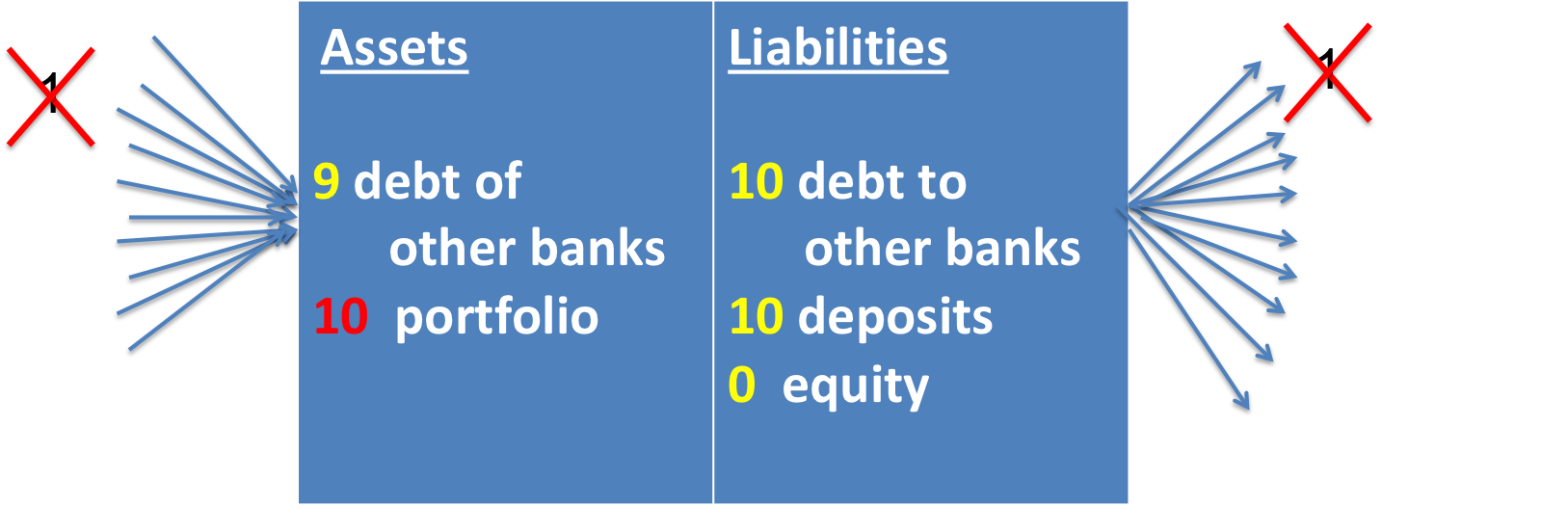}}
\caption{\small With correlated portfolios, banks are now more susceptible to defaults of others, even when levels of exposure to any single counterparty are low.
This can undo the benefits of diversification and the non-monotonicity discussed and now a second bank defaults even if it has ten partners.}\label{fig:nonmonotone5}
\end{figure}

Correlation in portfolios can mitigate and even erase this non-monotonicity in
contagion,
by removing gains from diversification and increasing
common vulnerabilities. To illustrate the role of correlation, reconsider the example
introduced above (Figure \ref{fig:nonmonotone3}).
Suppose banks' portfolios  exhibit substantial correlation -- perhaps because they all have
substantial exposure to the same sort of collateralized debt obligations, as was the case in 2007.
For example, suppose that when one bank's portfolio
drops below 10, the portfolios of other banks are also likely to be at below-normal levels. 
If we change the portfolio of the second bank pictured in Figure \ref{fig:nonmonotone3} (Right)
to drop to 10 at the same time as the first bank's portfolio dropped to 8, then
the default of the first bank is enough to push the second bank into insolvency---this is pictured in
Figure \ref{fig:nonmonotone5}.  In addition, the asset sides of banks' balance sheets are further depressed not
only due to the fact that their portfolios are weak at the same time, but because more than one of the debts that they are owed
are defaulted upon at once due to the correlation in other banks' values.\footnote{For other examples and simulations of the impact of
correlated portfolios
on systemic risk, see the online appendix of \cite*{elliottgj2014}.}

One way to understand this effect is that positive correlation in investments across
banks erases some of the benefits of diversification in counterparties, and facilitates contagion.
More generally, increasing the correlation in portfolios of investments leads to
increased probabilities of co-defaults.  For example, \cite{wagner2010} considers two banks and two assets,
and makes the observation that if both banks fully diversify their portfolios by equally dividing it
between the two assets, their portfolios end up being perfectly correlated.
Hence there exists a trade-off as more diversification -- which comes hand
in hand with more correlation between banks' portfolios -- reduces the unconditional probability
that each bank defaults, but can push up the probability that they default together (presuming
they were invested in different assets to begin with), hence increasing
systemic risk.  Of course, the worst case is when the banks have similar and under-diversified portfolios -- for instance,
all holding similar mortgages or loans -- as then they are correlated and risky.

\subsection{Robust Yet Fragile}\label{robustfragile}

As stated by \cite{haldane2009}, and studied in detail by \cite{gaik2010}, financial networks have an intriguing property of
being ``robust-yet-fragile''.\footnote{See also \cite{callawayetal2000} for an earlier discussion of percolation on graphs and such a tradeoff.}
Interdependencies between banks, in the form of lending or liquidity provision for instance,
allow for risk-sharing, which can help individual institutions be less susceptible to
individual liquidity or portfolio shocks.
Those shocks are spread among counterparties, and this sort of diversification helps lower the
chance of any individual institution's failure.  This is the sense in which financial networks are robust.
However, very large shocks can cause an institution to fail despite the diversification,
and then interdependencies can
transmit the shock more widely and more extensively.
There are, of course, nuances on this that depend on the model and the type of
contracts that exist between institutions (e.g., see
\cite*{alleng2000,gaik2010,galek2007,elliottgj2014,acemogluot2015}).

This  is related to the nonmonotonicity discussed above, but is a distinct phenomenon.
The robust-yet-fragile property is that one network can be an improvement over another for some situations,
but can also make things much worse in others.
In the above discussion of non-monotonicity, we considered how changes in the network affect whether or
not a particular shock cascades -
so the comparative static was in the network holding the shock constant.
The robust-yet-fragile phenomenon
is instead a comparison of how a network fares against different types of
shocks.

\cite*{acemogluot2015} focus on networks of unsecured interbank debt, and study how a shock to a
bank's returns propagates through the network. They distinguish between two shock regimes: shocks that are
small enough to be absorbed by total excess liquidity in the system, and those that are not.
In the former regime, interdependencies unambiguously alleviate the risk of contagion: the network
structure most resilient to contagion is the complete network, in which each bank's total liabilities are spread equally across all other banks. This leads to maximal risk sharing, and minimal expected number of defaults. However, if shocks are larger than the total excess liquidity in the system, interdependencies just facilitate its propagation.
One can see a foreshadowing of this point in the analysis of \cite{alleng2000} who show, in a more specific setting, that the risk of contagion depends on whether there is aggregate uncertainty about the demand for liquidity. If not, interbank connections increase risk-sharing without generating systemic risk.

 \cite*{cabralesgv2017} highlight the role of the size of shocks in a different model:
 one in which interdependencies between banks capture the correlation in their investments.
 They consider a set of ex ante identical banks, each having debt due to outsiders and
 access to a risky project. Returns to these projects are subject to shocks that are identically and
 independently distributed across banks. If a bank is unable to cover its debt to outsiders,
 it defaults and incurs some costs due to distress. Banks have the possibility of diversifying their
 portfolio by exchanging claims on each other's projects: the link from $i$ to $j$ captures the claim that
 bank $i$ has on the return of $j$'s project. Hence linkages in their model represent correlation in
 portfolios across banks, and not any sort of interbank obligation. The same trade-off emerges: more
 links allow for better risk sharing, but also entail being exposed to more sources of risk.
 The network structure leading to the least expected number of defaults depends on the distribution of
 shocks. In particular, they show that if large shocks are likely, then the empty network is optimal.
 If shocks are mostly small, then the complete symmetric network is the most resilient. Other shock distributions can lead intermediate network density to minimize contagion.

\paragraph{Heterogeneous Network Structures.}

\cite*{acemogluot2015} characterize the network structures most or least resilient
to contagion under different shock regimes.
If shocks are small, the complete network leads to the lowest risk of contagion.
In contrast, the ring network, in which  each bank's claim is concentrated on a single counterparty,
is the structure most prone to contagion.
The analysis applies to {regular} networks, in which all banks have as much interbank claims as
liabilities, and all have the same number of counterparties. This rules out heterogeneity in
bank size and in connectedness, as well as the possibility that a bank be a net lender or borrower.

Generally, analytic tractability limits full characterizations to such simple settings.
An alternative is to use properties of large random
networks.
Each technique offers valuable insights and tractability, but applies to limited classes of networks.

A challenge is that financial networks often
involve significant asymmetries -- such as the
presence of a core-periphery structure --
and that affects the risk of contagion.
Large core banks can be resistant to small shocks,
but can fail catastrophically when hit with large shocks, especially when those shocks are correlated.
This matters, as shown for instance in \cite*{elliottgj2014} who use simulations to document how core-periphery networks
can also erase some aspects of the non-monotonicity discussed in Section \ref{monotonicity},
since the failure of a large bank or entity to which core banks have large exposure can lead to
extensive contagion within the core,
and then spread to the whole system.
This was what loomed in 2008.
There are also other studies that provide evidence that heterogeneity makes a substantial difference.
For instance, \cite*{gai-haldane-kapadia} provide simulations showing that contagion varies
with the level of concentration in networks of interbank claims, \cite*{glassermany2015}
provide some theoretical results showing contagion is the largest when banks are heterogenous in
size and the shock originates at a large central bank, in a certain class of networks,
and \cite*{teteryatnikova2014systemic} shows how a negative correlation between neighboring banks'
degrees (numbers of counterparties) can help make the network more resilient.

Given the complexity involved it makes sense to continue analyses in two directions:
the development of more insights into how the structure of heterogeneity matters in financial contagion,
and specific applications to provide more empirical background on features of networks
that are prevalent and relevant.   Another direction is to apply models of contagion to observed networks for simulation of risk patterns, an approach increasingly being used by regulatory institutions (\cite{aikman2009funding, BCBS2015}).

\subsection{Self-Fulfilling Feedback Effects and Freezes: The Role of Cycles}

Next, we  discuss how network structures matter for the second type of systemic risk: self-fulfilling
feedbacks that generate multiple equilibria.

A burgeoning literature has highlighted the role of cycles in generating multiple equilibria
(see  \cite{rouknybs2018, jacksonp2020, derricor2019}), as was previewed in Figure \ref{feedback}.
In a network of interbank obligations, cycles enhance counterparty risk without
creating value in the financial system -- clearing cycles hence reduces the risk of default
cascades without impacting the values of banks.
For example, \cite{rouknybs2018} show that there exists multiple equilibria for bank values if and only if there is a
cycle composed of banks that are sufficiently interconnected,
such that a bank's solvency depends on the solvency of its predecessor in the cycle.

\begin{figure}[!h]
\begin{center}
\includegraphics[width=0.8\textwidth]{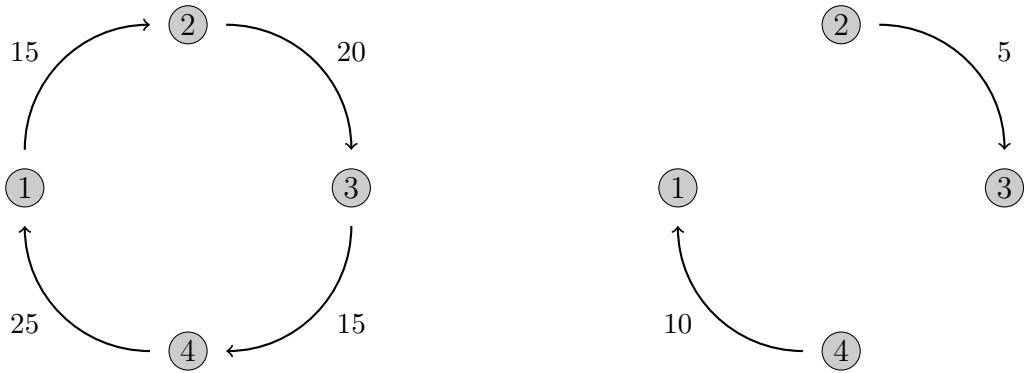}
\caption{\small 
Both networks have the same net debts, but one has a cycle and the other does not.  All banks have portfolios worth 10.   If a bank becomes insolvent, it pays none of its debt.    In the network on the right, there is a unique equilibrium in which all banks are solvent.  In the network on the left there are two
equilibria:  one in which all banks are insolvent and default, and the other in which no banks default.  }
\label{fig:cycles}
\end{center}
\end{figure}

As an illustration, consider the financial networks depicted in Figure \ref{fig:cycles}.
In the network on the left of Figure \ref{fig:cycles}, if no bank pays its debt, then given
that each bank's portfolio value of 10 is less than the debt it owes, no bank can afford to pay its debt.
This becomes self fulfilling: there is an equilibrium in which all banks are insolvent and no debts are
paid.    There is also an equilibrium in which all debts are paid.
In contrast, in the network on the right, the gross debts have been netted out and only net debts remain.
There is then a unique equilibrium in which all banks are solvent.
Clearing all cycles eliminates the possibility of multiple equilibria,
and ensures that only the best equilibrium
remains.  Thus, there can be large gains from clearing cycles,
especially if one expects banks, and
investors more broadly, to hold pessimistic beliefs - or believe that others hold pessimistic beliefs.

Of course, this example is stylized to make it clear how banks could net out their debts and
avoid the bad equilibrium.  However, in practice these cycles are much more complex,
involving many banks, different debt
maturities, and other factors that can lead to inertia and make such a coordination failure more likely,
especially if times are uncertain and people delay payments on debts.
Thus, it is important to understand when cycles can be
problematic and how to overcome them when they exist.

\cite{jacksonp2020} study self-fulfilling cascades and freezes in detail and characterize
how they rely on the presence of cycles,
the portfolios of the banks, and costs and delays involved in insolvencies.%
\footnote{In the absence of bankruptcy costs, bank values are generically
unique (\cite{eisenbergn2001}).}
Such self-fulfilling cascades and freezes become less likely if some banks have lower gross exposures --
as these banks may then have enough capital to act as buffers and stop the cascade.
Self-fulfilling cascades occur whenever no bank on a given cycle has enough assets coming from
outside the cycle to spark a repayment cascade in that cycle.
\cite{jacksonp2020} also allow for more general financial contracts between banks, and
characterize the existence of multiple equilibria, showing that
they are generated by a particular type of cycle involving some debts, but potentially other contracts too.
But they show that cyclical structures that do not involve some key debt obligations cannot lead to
multiple equilibria.
This characterization of cyclical structures of a network not only explains
equilibrium multiplicity and freezes, but also provides the basis for
identifying the minimum bailouts needed to return
a network to full solvency, as discussed below.

Gains associated with clearing cycles can provide some insight into the use of a risk management technique
called portfolio compression (\cite*{derricor2019}). This technique consists in a netting mechanism
that aims at reducing gross interbank exposures, and hence
regulatory requirements.  A similar intuition is often put forward when discussing the benefits from
clearing bilateral OTC trades through central counterparties (CCPs). CCPs allow for multilateral
netting of interbank contracts, which enhances transparency of the financial network and limits
counterparty risk.\footnote{See \cite{duffiez2011} for more detailed discussion and background,
and \cite*{capponi2018clearinghouse, wang2020theory} for more recent papers on the design of
margin and collateral requirements for CCPs.}
If interbank contracts are restricted to debt contracts, such multilateral netting boils down to clearing cycles, as done in Figure \ref{fig:cycles}.


\section{Incentives in Financial Networks}\label{incentives}

Systemic risk depends on several factors -- including network structure, the portfolios of  institutions and their correlation --
that are endogenous: these result from choices by the institutions that compose the network.
Thus, it is vital  to understand whether institutions have efficient
incentives; i.e., incentives to make investments and choose partnerships that maximize the overall
value of the financial system.

Given that financial networks are full of externalities, we should expect individual financial incentives to fail to align
with the overall welfare of the economy.
Indeed, the literature shows that incentives are misaligned on many dimensions.
We start by reviewing how interconnections between banks, and the potential for contagion they induce, affect banks' investment decisions in \emph{outside} assets.
We then look at banks' incentives when choosing their \emph{interbank} assets,
and how these impact the equilibrium network structure.

\subsection{Investment Decisions}

The literature has highlighted two main distortions in banks' investment decisions: they have an incentive to take on too much risk, and to correlate their portfolios with that of their counterparties. We discuss how the network structure comes into play in both of these inefficiencies.

\paragraph{Inefficiently Risky Investments.} There are two main ways in which financial interdependencies can lead banks to take on too much risk compared to what is socially  optimal, and they relate to the two types of systemic risk identified in Section \ref{types_risk}.

First, a bank's investment decisions not only impact its own value, but also indirectly the values of its
counterparties, and of its counterparties' counterparties, and so on.
This sort of externality is not new to the financial network setting: there are many settings in
which the choice of investments might not
reflect the interests of all those who are impacted (\cite*{admatih2013}), with an early illustration of this
being made by \cite{jensenm1976} in which a manager
makes choices that do not reflect shareholders' interests.
This has been investigated in a variety of network settings
(\cite*{bruscoc2007,hirshleifert2009,galeottig2019,jacksonp2019,shu2019}), where the externalities are very wide
and the interests extend well beyond those directly interacting with an institution.

As an illustration,
 \cite{jacksonp2019} consider the following
 portfolio choice problem, in which each bank $i$ has access to a safe asset with constant rate of
 return $1+r$ and a risky asset with random return $\tilde{p}_i$, with $\mathbb{E}[p_i]>1+r$.
 Banks are furthermore linked to each other via financial contracts, that either take the form of
 debt or equity. Fully investing in the risky asset is a strictly dominant strategy for a bank as
 soon as the bank does not belong to a certain type of cycle in the network; even though this
 can result in excess systemic risk.
 Intuitively,
 only specific cycles generate the possibility that the bank's risky behavior may feedback to
 itself through the network by triggering a default cascade. Without the risk of such feedback,
 banks overlook any external costs they trigger when weighting the benefits and costs of a riskier
 investment, leading them to over-invest in the risky asset.

Second, network interdependencies can make a variety of banks' decisions strategic complements,
leading to socially inefficient outcomes.
For example, \cite*{allouch2017strategic} consider a network of interbank liabilities, in which banks have the possibility of storing part of their initial cashflow to overcome any future net deficit, instead of cashing out these benefits right away at the cost of foregone future returns.
They show that this can be viewed as a coordination game in which banks choose either Default or No Default.
Banks' decisions are  strategic complements, since it is easier for a bank to remain solvent if other banks
remain solvent as well. They show that, if there are cycles of debt in the network,
there can exist bad equilibria in which banks choose to cash out early and then default because they
expect others to do so as well. These equilibria are inefficient as all banks, as well as their outside
creditors, would have been better off had they all coordinated on remaining solvent.

\paragraph{Endogenous Correlation of Investments.}
There are many forces that push financial institutions to correlate their investments.

Some are basic herding forces:  seeing others make an investment may signal something about that investment's prospects (\cite*{bikhchandanihw1992,banerjee1992,chincarini2012}), or the people who are choosing investments may worry about their reputations (\cite{scharfsteins1990}).  Other forces pushing towards correlation  are regulations that limit the scope of investments, essentially pushing them to hold certain classes of assets, minimum amounts of certain assets, or to have a portfolio that meets certain risk characteristics.  Banks also have forces that push them to the same lending strategies as their competitors (e.g., \cite*{cohen2015static}).
Though the aim of such regulations is to make portfolios safer, the
fact that they push banks to hold similar assets can make
rare negative shocks (e.g., a major sovereign default) hit many institutions at the same time.

Beyond these forces, there are also network forces.
A first aspect of importance is the incentives of the regulator when
deciding whether to bail-out insolvent banks. If more banks fail at once,
a default cascade is more likely to be triggered, and the regulator has
more incentives to step in. \cite{acharya2007} highlight this
``too-many-to-fail'' problem that incentivizes banks to correlate their
portfolios, because if they all become insolvent together then they are all
more likely to be rescued.
Another driving force of banks' incentive to herd is the risk of information contagion,
which we described in Section \ref{types_risk}.
In a world of incomplete information, adverse information on some banks can increase borrowing
costs of others, because investors negatively update about the creditworthiness of the latter.
Such inferences make it less valuable to have an uncorrelated portfolio  and,
as \cite{acharya2008information} show, banks prefer to have correlated investments.
Yet another force is detailed by
\cite{elliottgh2018}, who consider a model in which banks can swap claims
on each others' investments, in order to hedge shocks to their exposures.
There are no interbank liabilities, but each bank owes some debt to outside
investors. Because shareholders act under limited liability, they have an
incentive to shift losses from them to debt holders. Such risk-shifting
motivates banks to correlate their portfolio returns, such that if one is
hit by a large shock, so are others, they all default, and losses are born by creditors.
Finally, in a model that accommodates debt and equity claims between banks, \cite{jacksonp2019} show how incentives derive from counterparties' portfolios.
All else equal, a bank prefers to be solvent when it earns the most returns from its contracts with other banks.   This pushes it to prefer to be solvent when others are solvent, and insolvent when others are insolvent.   This leads
perfect correlation of portfolios across counterparties to be an equilibrium, and in fact the Pareto-dominant equilibrium from the banks' shareholders' perspective.
\bigbreak

Private party
monitoring and reputations can also help mitigate incentive issues, as
poor investments raise the
capital costs of financial institutions (\cite*{godlewski2012bank,godlewski2018financial}).
This can affect banks' choices of portfolios and partners in the network.
Although, dynamics can help mitigate some problems,
moral hazard problems are generally not extinguished by reputations and
monitoring (e.g., \cite{diamond1991monitoring,rajan1992insiders,holmstrom1997financial}).
How banks' incentives are impacted by such feedbacks in financial networks is an
important open topic.

\subsection{Incentives in Network Formation}

We now review what is known about banks' incentives when choosing their
counterparties in the financial network, and highlight inefficiencies.
Inefficient network formation is a recurring theme in the literature, starting
with \cite*{jacksonw1996}, but plays out in particular ways in the context of financial institutions.

Several factors can lead a core-periphery structure -- empirically observed in many financial markets -- to arise
endogenously, some of which are
based on the various roles of
core banks as intermediaries in financial markets.
\cite*{babush2017} show that when partnerships between banks not only have a trading function
but also an informational function then there are economies of scale in intermediation and
the equilibrium network has a star structure.
The central agent has information about all others and enforces all contracts, and ends up intermediating all trades at some fee. Because the star network leads all relevant information to be centralized, it is constrained efficient.
In \cite{farboodi2017}, it is not information frictions that drive intermediation,
but banks' unequal access to investment opportunities. Banks that have access to these
opportunities constitute the core, and funds
flow to them from the periphery.
The equilibrium network is socially inefficient --
core banks over-connect whereas periphery banks under-connect --
and the core captures intermediation rents.
Finally, as \cite{wang2017} shows, a core-periphery structure can also stem from inventory efficiency.
In a market in which institutions have random trading needs, intermediaries can arise to complete those trades.   This leads to
random inventories for the intermediaries,  and thus concentrated intermediation reduces inventory risk via a law of large numbers.
This trades off against market power of intermediaries.  As Wang shows, this leads to a socially inefficient equilibrium network, with either too few or too many dealers (core banks) depending on the asset's trading frequency.

The potential inefficiency of a core-periphery structure usually comes from banks choosing either too few,
or too many, counterparties given the externalities
arising from, and the market imperfections that drive, intermediation.
Beyond core-periphery structures, the number of counterparties a bank chooses can
generate inefficiencies in itself, as it impacts systemic risk,
which is less than fully internalized by banks.
\cite{acemoglu2015systemic} show that banks tend to lend too much to each other,
and spread their lending insufficiently across borrowing banks. This under-diversification of
interbank liabilities appears despite the fact that equilibrium interest rates reflect the risk-taking
behavior of borrowers: bilateral externalities are internalized via equilibrium interest rates,
but not the more general network externality, leading the equilibrium network to be socially
inefficient.
Similarly, an analysis in \cite{jacksonp2019} shows that
banks tend to choose too few counterparties on which to hold claims,
because they overlook contagion costs others incur when they go bankrupt.\footnote{The incentives
can also depend on the structure of the contract and
bargaining between counterparties, as shown by \cite{duffiew2016}.}

Furthermore, the equilibrium network  exhibits
`homophily' in portfolios of outside assets (\cite*{elliottgh2018}).
Banks prefer to be counterparties of other banks that have similar portfolios.
This is the flip side of choosing portfolios that are correlated with those of one's counterparties.
When choosing on whom to hold claims, banks acting under limited liability under-value diversification
and prefer to be linked to those with similar portfolios. This implies that, when a shock hits,
banks default together and shift losses to debt-holders.

Inefficiency is not the only prediction of the literature on financial network formation.
Some papers document forces that push to the formation of networks that avoid widespread contagion. In particular, \cite{babus2016} shows that if banks can commit to mutually insure each other against liquidity shocks, then there exist equilibria in which contagion never occurs.
\cite{erolv2018} consider networks that are stable to deviations by groups of banks who can restructure their connections
and sever outside ones.   In their model, a bank defaults as soon as one of its counterparties
does, so if any bank in a component defaults, then all will.   Given a positive benefit from forming a
connection, each bank should connect to all other banks that it already has a path to.
This leads to the prediction that a stable network will be a collection of
fully-connected disjoint clusters.   Group stability ensures that the number of banks in each
cluster will be the one that maximizes their overall value.
Although empirically observed networks have much more heterogeneity,
the reasoning behind their results can help explain why the `core'
in core-periphery networks is often fully connected.

Finally, fear of contagion can lead to credit freezes, whereby banks abstain from lending to each
other, resulting in an overly sparse network.
\cite*{acemoglu2020systemic} discuss how the extent of a credit freeze depends on the structure of the underlying network of potential partnerships between banks and the distribution of liquidity shocks. They show how the financial system may fail to allocate capital efficiently from depositors to entrepreneurs due to intermediation frictions.


\section{Regulation and Intervention}

We now turn to issues surrounding regulation by a government
authority that is interested in overall societal welfare, and understands
the systemic risks and inefficiencies discussed above.

\subsection{The Necessity of Network Information}

As should be obvious by now, properly addressing systemic risk involves a holistic view of the
network.\footnote{Attempting to assess systemic risk without detailed network
information is what \cite{jackson2019} refers to
as ``flying jets without instruments'':  operating a complex interactive system without
the necessary measurements.  Even though some measures
 that work without network information (e.g., S-risk) may correlate with more precise
full network measures, if they are only
approximately capturing the real risks, they can be ineffective.}
The following example illustrates the importance of seeing details of the network in order to assess which
are the key institutions to regulate and/or  bailout.

An important component of systemic risk assessment is stress testing, which is usually run in a decentralized manner.
The main input into many stress tests is balance sheet data, which describes the amount of each type
of financial assets and liabilities held by each bank. Depending on the jurisdiction, balance sheet data does not always provide
complete, or even partial, information about one's counterparties,
and hence about the network structure.
Such ``local'' data can completely miss which banks are most likely to start a default cascade,
or be caught up in one.
The point is straightforward, but worth emphasizing given its importance.

\begin{figure}[!h]
\centering
\includegraphics[width=0.45\textwidth]{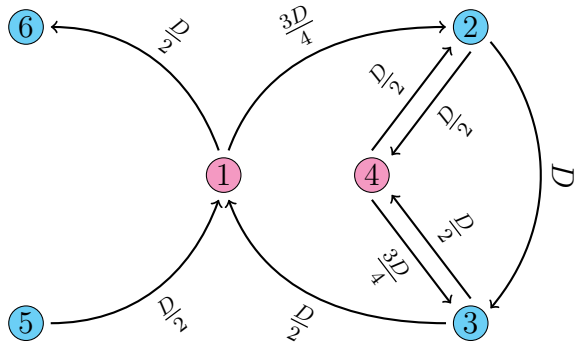}
\caption{Arrows point in the direction that a debt is owed. Banks 1 and 4 (magenta) have the same
balance sheet: they both have two debt claims on other banks (both of $D/2$), and two
debt liabilities (one of $D/2$ and one of $3D/4$). They are both net debtors.}\label{fig:contracts}
\end{figure}

For ease of illustration, consider a network in which banks are only linked via debt contracts.
A measure of systemic risk based on local balance sheet information only depends on the face value of each bank's assets and liabilities, but not on the {identities} of its counterparties.
To show why this is insufficient information, we give an example of financial network in which two banks
have {identical} balance sheets, and yet their defaults have significantly different consequences.
Hence if the central authority were able to bailout one (and only one) of the two institutions,
it could not make an optimal decision based on local information. Consider the network
depicted in Figure \ref{fig:contracts}.

Suppose the portfolios of Bank 1 and 4 both yield $0$, so that they are both insolvent.
Let Bank 2 earn a return on its portfolio that falls between $3D/4$ and $D$, Bank 3 below $D/4$,
and Bank 5 above $D/2$. Let the recovery rate on assets of a defaulting bank be zero.
Despite the fact that Banks 1 and 4 have the same balance sheet, only the former induces
widespread default contagion if it remains insolvent. Indeed, Banks 2 and 3 have enough buffer to absorb
the shock of Bank 4's default,
but not that of Bank 1. Hence, bailing out Bank 1 prevents the whole system
from insolvency, while bailing out Bank 4 does not change anything and a full systemic failure occurs.

This example also highlights the fact that, without network information,
one cannot even identify which banks are at risk of insolvency.
For instance, if one examines the books of Bank 3 without knowing that Bank 2
is exposed to Bank 1, even if one knows the portfolio realizations of 3's
counterparties, but does not know the looming failure of Bank 1, it will appear that Bank 3 is
free from danger of insolvency.

While this is a simple example, it illustrates why regulatory agencies that cannot see
parts of the network (e.g., foreign institutions, shadow banks, etc.)
or have only data from local stress tests,  are at a substantial disadvantage.

Accordingly, and partly spurred by the lessons learned from the 2008 financial crisis,
 assessments of systemic risk that involve nontrivial portions of the network are beginning to emerge.  For example, the European Central Bank has information on the counterparties involved in the largest exposures of most banks within its jurisdiction.
This permits the construction of a network of a portion of the assets and liabilities within the
European banking sector, and some pointers to banks outside of Europe.  Accordingly, some
calculations of systemic risk of a nontrivial part of the network are beginning to
emerge (e.g., see \cite*{covigk2018,farmer2020foundations}). Similarly, the Bank of England has regulatory data on bilateral transactions between UK banks, allowing for the analysis of the UK interbank network in different asset classes (see \cite{ferrara2017, bardoscia2018}).
This is an important advance in the assessment of systemic risk, but much more is needed and especially outside of Europe and for the growing shadow banking system
which falls outside of most jurisdictions.

\subsection{Addressing Systemic Risk}

With some network information in hand, and an understanding of the issues discussed above, we
can think about
optimal interventions in financial markets.
We discuss this in two parts:  one about avoiding cascades,
and the other about avoiding bad
equilibria and feedbacks in settings with multiple potential outcomes.  Let us start with the second one first.

\paragraph{Eliminating Self-Fulfilling Feedbacks.}
The literature has brought forward several ways a regulator can intervene to
address  systemic risk stemming from multiple equilibria and self-fulfilling
crises. Such crises are the consequence of a coordination problem between
some agents taking part in financial markets. For instance a bank run arises
when depositors mis-coordinate and all withdraw their deposits, and a credit
freeze when banks all choose to abstain from lending. Importantly, in both of
these cases, there exists another equilibrium in which no crisis occurs, and
which is preferred by all. One way to eliminate this source of systemic
fragility is to have the regulator insure all lending, either via deposit
insurance (\cite{diamondd1983}) or by committing to be the lender of last
resort and inject capital whenever necessary (\cite{diamondr2011,
bebchuk2011self}). If credible, this eliminates the possibility of
miscoordination.

Self-fulfilling crises due to coordination failures differ from contagions triggered
by fundamental losses in that self-fulfilling crises can be stopped at no capital cost to the regulator.
This is a key distinction between the two types of systemic risk. By guaranteeing deposits or
lending,  a regulator ensures that no bank run or credit freeze will lead banks to default, and
hence will not have to intervene in equilibrium.

Even if a regulator does not insure interbank lending ex ante, and a
self-fulfilling default cascade or freeze occurs (Figure \ref{feedback}),
it can still be ended by injecting (at least partially recoverable) capital in the network
appropriately. \cite{jacksonp2020} show that such cascades stem from the
presence of cycles of claims in the network, and that stopping them requires
injecting enough capital to clear these cycles.
The injection of capital is a way of ``jump-starting'' the payment cycle, avoiding a bad equilibrium.
They also show that any capital injected into self-fulfilling cycles can be fully recouped by the regulator,
making this part of the bailout policy virtually costless.
To illustrate this, reconsider  the network on the left of Figure \ref{fig:cycles}.
If someone were to inject 5 units of capital into either Bank 1 or 3,  then there is a unique equilibrium in which all banks are solvent:  1 can pay 2 (even without any inflow from 4), which then can pay 3, etc.
Moreover, that capital can then be recovered by the injecting authority once all debts
are paid. 
In contrast, note that injecting 5 units of capital into Banks 2 or 4 would not have the same effect:  the bad `all default' equilibrium would still exist.
More generally, as \cite{jacksonp2020} show, there are key banks that are most advantageous to inject capital into in order to
ensure that only the best equilibrium remains, and this depends on the leverage that
their payments provide in the network.
This can also depend on which banks lie on multiple cycles at once,
which happens in more complex networks.
They characterize the minimal injections of capital needed to
restore solvency, and show that finding the least expensive
approach is a complex (NP-hard)
problem (when many banks are involved).  However, they also show that in some well-structured networks, such as core-periphery networks with some
symmetries in the sizes of banks and balance sheets within the core, finding optimal bailout policies
is more straightforward and intuitive.

Another solution that has been brought forward to avoid miscoordination and self-fulfilling
crises is to have all transactions go through a Central Clearing Counterparty (CCP).
Centralization helps because it allows for multilateral nettings of obligations, which
in particular eliminates cycles of claims and reduces the possibility of multiple equilibria.
\cite{csokah2018}  highlight the gains from centralization when clearing payments between banks.
They show that a centralized clearing process always yields the best equilibrium for bank values,
whereas a decentralized process converges to the worst equilibrium.

Of course, these interventions may distort banks' investment incentives ex ante,
leading them to take on even more risks, and be socially costly in that sense --- we discuss the
interplay between regulation and incentives in Section \ref{regulationfeedback}.

\paragraph{Ex Ante Reserves or Capital Requirements Versus Ex Post Bailouts.}
As discussed in Section \ref{section_contagion}, an initial loss by some financial institution
 can get amplified by network interdependencies and spread through the financial system.   The ensuing bankruptcy costs are real losses to the economy,
 and can be avoided, or at least minimized, by intervention.

There are two main ways in which a regulator can intervene that have been considered.
One is to regulate banks ex ante so as to ensure that inefficiently risky investments are avoided.
 This can be done via various forms of prudential regulation, for instance, by imposing reserve, liquidity, or capital requirements,\footnote{See \cite{martinezetal2020} for a discussion
 of the differences between various capital and liquidity requirements and how their effects depend on bank size and business cycles.
 Also, such policies interact across jurisdictions and, for instance, \cite{karamyshevas2020} show
 that prudential policies have substantial spillovers in risk reduction across countries.}
 constraining the types of investments that different institutions can make,
 and monitoring banks' capital ratios and
 investments on an ad hoc basis.
 A second is to allow arbitrary investments but then intervene
 and  inject capital if some
 danger of cascading defaults arises, in order to minimize
 contagion.    As shown by \cite{jacksonp2019},   whether one wants to
 intervene depends on the financial centrality (more on that below)
 of the bank in question.\footnote{See also \cite{belhaj2020prudential} for more discussion of centrality and prudential regulation.}
 If a bank is sufficiently central
 so that it poses substantial systemic risk, then whether it is better to
 regulate it ex ante or bail it out ex post,  depends on the relative opportunity costs of the
 excess returns lost by forcing the bank to hold safer assets compared to
 the real costs of a bailout.\footnote{See \cite{lucas2019measuring} for an estimate of bailout costs in the 2008 financial crisis.}$^,$\footnote{ There are further issues to be considered.   For example, when bailing out a bank, one can do so
 by providing capital with some hopes of that being repaid in the future.
 Providing that capital in the form of debt can end up just changing the timing of the default,  while offering bailout money
 in an equity form avoids imposing additional constraints on the payments the distressed institution has to make.}

The determination of which financial institutions contribute the most to systemic risk, can be measured via notions of financial centrality.
 The literature has suggested several  measures of  centrality,
 aiming to assess either the exposure of a given bank to systemic risk or
 how much it itself contributes to it.\footnote{Paying attention to centrality
 makes more of a  difference in asymmetric networks, such as core-periphery ones,
 compared to more regular networks (e.g., see \cite{capponi2015systemic}).}
 Some of these measures are
 solely based on market data on portfolio returns of individual banks, and
 include features of contagion either through fire sales
 (\cite{duarte2018fire, engle2019measuring}) or the correlation structure of returns between
 bank portfolios (\cite{billioetal2012}). Others are based on the
 underlying network of interbank contracts and rely on models of contagion via
 counterparty risk (\cite{amini2016resilience, hauton2016,jacksonp2019}).
For example,
\cite{jacksonp2019} propose a measure of a bank's ``financial centrality'' based on its systemic impact when fundamental asset prices go from $p$ to $p'$.
Given a network of interbank contracts and liabilities, one can trace how this change will cascade through the network and
affect the solvencies and values of all institutions.    By seeing how the change in a given bank's portfolio  cascades, one can assess its systemic importance.
This measures the eventual total change in the value accruing to all outside investors in the financial system.

From another perspective,
\cite{demange2016} proposes a measure of spillover effects in a network of
interbank liabilities that relies on the properties of equilibrium debt
payments between banks. She defines an institution's threat index as the
marginal impact of an increase in its direct asset holdings on total debt
repayments in the system. This index is null for all institutions that are solvent, as they are already able to pay back their debt in full. It is strictly positive for defaulting institutions, as a larger portfolio means that they can repay a larger fraction of their liabilities, which may enable other banks to repay more of their liabilities, etc. The extent to which this spreads through the network is then captured by the institution's threat index.   This can be viewed as a marginal version of the measure above,  where marginal refers to changes that are small enough not to change any of the defaults, but just the payment streams.

 Public bailouts come at a cost to the regulator, and in particular can depress the price of government bonds when they are financed by debt. If banks hold large amounts of sovereign debt, this further worsens the value of their portfolio, and even larger bailouts are required to maintain their solvency. This ``doom loop'' was a key contributor to the sovereign debt crisis in Europe following the Great Recession. \cite*{capponi2020optimal} propose a model incorporating this amplification mechanism, and characterize optimal bailouts in this setting.

 All these papers focus on networks of financial organizations,
 but a regulator should be interested in the impact on the overall economy.
 Contagion is indeed not specific to financial markets: for instance the structure of input-output
 networks and supply chains can induce small shocks to magnify in a similar manner as a financial
 network can amplify shocks to returns and affect asset prices
 (e.g., \cite*{acemoglucot2012, barrot2016input, ramirez2017, herskovic2018networks}).
 Determining the optimal injection of capital then requires a good understanding of the interplay between financial networks and supply chains, which is a topic that has not yet received a lot of attention.

\subsection{Feedback between Regulation and Markets}\label{regulationfeedback}

The previous section considered interventions to address systemic risk, but took as given the network and
portfolios.  Of course, these are endogenous, and changes in regulation can affect them.
This interplay between regulation and incentives of market participants is only partly understood.
Accounting for responses to regulation can change some comparative statics and policy implications.

\paragraph{Regulation and Investment Incentives.}
Regulation that aims at reducing systemic risk can have perverse effects on banks' investment incentives.

For example, bailouts, if anticipated, can lead to moral hazard as banks then no longer suffer the costs of risky investments.%
\footnote{This can even happen with private bailouts, as shown by \cite*{elliottgj2014} (see the supplemental appendix).
A bank that knows it will be bailed out by another bank in the event of
insolvency can have incentives to make riskier choices as it bears less of the consequences of those insolvencies.}
The mechanism through which this works is nuanced, since the shareholders who control a bank may not gain value during
a bailout, whereas debt-holders would.  But there are a variety of
reasons why the management of a bank may wish to avoid insolvency, that
have nothing to do with equity value.
There is evidence that this matters as, for instance,
\cite{damk2012}  find that bailouts
led German banks to take on more risk.
 Similarly, \cite{calomirisj2019} look at the effect of the introduction of deposit insurance in the U.S. in the early 19th century. They find that deposit insurance reduced market discipline, and led to more risk taking on the part of banks. Hence reducing liquidity risk came at the cost of a higher risk of insolvency, because of the induced distortion of banks' incentives. Public bailouts not only affect banks' investment decisions, but also their choice of counterparty, and hence the equilibrium structure of the financial network (\cite*{erol2019}). Without bailouts, a network with high clustering and low concentration endogenously arises in response to the possibility of counterparty (and more importantly second-order counterparty) risk. The anticipation of bailouts however makes banks less concerned about contagion risk, introducing a ``network hazard'' that leads to low clustering and high concentration in the equilibrium network. This perverse effect of bailouts adds to the more standard moral hazard problem they induce.

At present, there is little empirical work on how the structure of the
financial network responds to changes in regulation.  A notable exception
is a recent study by \cite{anderson2019bank} that looks at the effect of the
National Banking Acts of 1863-1864 on the topology of the U.S. banking system.
The National Banking Acts established reserve and capital requirements, and created a reserve hierarchy. This led the banking system to become more concentrated around a few core banks, located in designated reserve cities.
This allowed for better diversification, but increased the potential for contagion if one of the core banks were to fail.

\paragraph{Public Bailouts, Private Bail-Ins, and Counterparty Choice.}
Given the high financial cost of public bailouts and the perverse incentives that they generate,
attention has been paid to private-sector resolutions of defaults, especially since the last
financial crisis.
Private-sector bail-ins -- private banks rescuing each other when
insolvent -- can be incentivized, and such incentives depend on the topology
of the financial network.
It can be in one bank's interest to rescue another if the gains from preventing its default are higher than the costs of rescue.
Gains come from the value of interbank assets, which are enhanced if a default cascade is prevented.
This means that linkages between banks not only spread shocks but also incentivize private sector bailouts, so that a more interconnected financial system can reduce systemic risk, and lead to more investments (\cite{leitner2005}).
\cite{kanik2018} shows how bankruptcy costs induce private-sector bailouts, as they magnify losses due to defaults.\footnote{In settings without any bankruptcy costs, incentives can completely
disappear, as shown by
\cite{rogers2013failure}. }
He examines the incentives of coalitions acting together to avoid defaults,
and shows that non-clustered networks with intermediate levels of interdependencies lead to optimal incentives.
Furthermore, if the network is clustered, then potential losses are not fully internalized by solvent banks,
leading to inefficient rescue levels.

The incentives can be complex, as many different entities all gain from avoiding defaults.
For example,
\cite{bernard2017bail} analyze the interplay between public bailouts and
private bail-ins.
Bail-ins can only be incentivized if the regulator can credibly commit to
not step in, which is only the case when contagion in the absence of
intervention is limited enough. For large enough shocks, interconnectedness
makes it in the regulator's interest to step in if no one else does, and
then bail-ins cannot be incentivized.  Furthermore,
banks contribute to a bail-in if and only if they get a high enough share of the induced gains,
which generally diffuse through the entire financial system. Hence, the more diversified a network is,
the less individual banks are willing to step-in and rescue each other.

\paragraph{Shadow Banking and Moving Targets.}
One important and problematic characteristic of financial networks is their complexity, which can make
them opaque.\footnote{In addition, the financial organizations involved may be concerned about keeping information about their trading positions and partners private.
See \cite*{hastings2020privacy} for a discussion of some related issues, and new methods of obtaining critical network information while preserving privacy.}
Financial markets  involve many different types of market participants,
who are trading various kinds of financial assets leading to complex and multidimensional interconnections.
Moreover, part of this complexity arises in response to regulation,
with financial innovations and new products being introduced to circumvent regulatory
restrictions (\cite{silber1983process}).\footnote{See \cite*{anderson2020interbank} for an analysis showing how growth in shadow banking interacts with interventions within the banking sector, and
can exacerbate risk.}
Importantly, a significant portion of these trades are realized in over-the-counter markets,
and are hence may only known to the two parties directly involved in the exchange, when not required to be
disclosed.
In the absence of full network data, one can design policies that are optimal in the face of uncertainty
about the structure of the network and costs of improving transparency, which is a new direction explored
by \cite{ramirez2019}.

Even if one has detailed accounting information from all of the financial institutions
within some regulatory jurisdiction,  the difficulty of
monitoring the financial network is aggravated by the increasingly large portion of the financial system that operates outside the jurisdiction of financial regulation, making it even harder for regulators to have a complete picture of the market.
In addition, the shadow banking system is endogenous, and can expand in response to stricter restrictions in the regulated system.
This was for instance one of the side effects of Regulation Q of the 1933 Glass-Steagall Act, which prevented banks from paying interests on checking accounts. When interest rates increased in the 1950s, it left room for the emergence of substitute forms of demand deposits that paid interest -- e.g. Savings and Loans, and money market deposit accounts -- leading investors to move their capital to the shadow banking system (\cite{lucas2013}).   Once those were regulated, one could see new forms of unregulated institutions emerge, and the variety of institutions that are involved in some sort of financial intermediation
is now quite enormous.

Taking a step back, it is far from clear what the scope of ``financial'' regulation should even be. Where should the regulator draw the line between institutions that are are considered as financial ones, and are regulated as such, and those that are not? Take for instance large corporations, such as private universities with large endowments. They are often both borrowing and lending at the same time, interacting with both the financial system and the real economy. In a similar manner as more traditional banks, they can spread shocks and take part in financial contagion  -- should they then be regulated in a similar way?

\subsection{Political Challenges}

As should be clear from our discussion above,  regulation is far from one-size-fits-all, and optimal intervention depends on many factors, including
network positions and centralities that are constantly shifting.    Unfortunately,  regulations are slow to adjust and often constrained by politics, since intervention can
benefit some parties more than others.
Historically, regulation has surged after financial crises (e.g., the Glass Steagall Act and Dodd-Frank)
and then slowly eroded over time until another crisis
hit.\footnote{See \cite*{aikman2019would, duffie2019prone, tarullo2019financial,jackson2019} for narratives of the
last financial crisis, and discussions of the regulatory framework pre- and post-crisis. }
Discretion granted to central banks and other regulators is one way of avoiding political cycles, but even that discretion changes with time.
In fact, that discretion can often be unclear, as was true in the 2008 crisis during which it was not obvious how much authority the Treasury and Federal Reserve had to intervene directly.    Building a regulatory system that quickly adjusts to constantly shifting financial networks is yet another challenge.

\bibliographystyle{ar-style1}
\bibliography{financeNetworks}

\begin{thebibliography}{}
\expandafter\ifx\csname natexlab\endcsname\relax\def\natexlab#1{#1}\fi

\bibitem[Acemoglu et~al.(2012)Acemoglu, Carvalho, Ozdaglar \&
  Tahbaz-Salehi]{acemoglucot2012}
Acemoglu D, Carvalho VM, Ozdaglar A, Tahbaz-Salehi A. 2012.
The network origins of aggregate fluctuations.
\textit{Econometrica} 80:1977–2016

\bibitem[Acemoglu et~al.(2020)Acemoglu, Ozdaglar, Siderius \&
  Tahbaz-Salehi]{acemoglu2020systemic}
Acemoglu D, Ozdaglar A, Siderius J, Tahbaz-Salehi A. 2020.
Systemic credit freezes in financial lending networks.
Tech. rep., National Bureau of Economic Research

\bibitem[Acemoglu et~al.(2015{\natexlab{a}})Acemoglu, Ozdaglar \&
  Tahbaz-Salehi]{acemogluot2015}
Acemoglu D, Ozdaglar A, Tahbaz-Salehi A. 2015{\natexlab{a}}.
Systemic risk and stability in financial networks.
\textit{The American Economic Review} 105:564--608

\bibitem[Acemoglu et~al.(2015{\natexlab{b}})Acemoglu, Ozdaglar \&
  Tahbaz-Salehi]{acemoglu2015systemic}
Acemoglu D, Ozdaglar AE, Tahbaz-Salehi A. 2015{\natexlab{b}}.
Systemic risk in endogenous financial networks.
\textit{Columbia Business School Research Paper}

\bibitem[Acharya et~al.(2007)Acharya, Bharath \& Srinivasan]{acharyabs2007}
Acharya VV, Bharath ST, Srinivasan A. 2007.
Does industry-wide distress affect defaulted firms? evidence from creditor
  recoveries.
\textit{Journal of Financial Economics} 85:787--821

\bibitem[Acharya \& Yorulmazer(2007)]{acharya2007}
Acharya VV, Yorulmazer T. 2007.
Too many to fail?an analysis of time-inconsistency in bank closure policies.
\textit{Journal of financial intermediation} 16:1--31

\bibitem[Acharya \& Yorulmazer(2008)]{acharya2008information}
Acharya VV, Yorulmazer T. 2008.
Information contagion and bank herding.
\textit{Journal of money, credit and Banking} 40:215--231

\bibitem[Admati \& Hellwig(2013)]{admatih2013}
Admati AR, Hellwig MF. 2013.
The bankers’ new clothes: What’s wrong with banking and what to do about
  it.
Princeton University Press: Princeton

\bibitem[Aikman et~al.(2009)Aikman, Alessandri, Eklund, Gai, Kapadia
  et~al.]{aikman2009funding}
Aikman D, Alessandri P, Eklund B, Gai P, Kapadia S, et~al. 2009.
Funding liquidity risk in a quantitative model of systemic stability.
\textit{Bank of England working paper}

\bibitem[Aikman et~al.(2019)Aikman, Bridges, Kashyap \&
  Siegert]{aikman2019would}
Aikman D, Bridges J, Kashyap A, Siegert C. 2019.
Would macroprudential regulation have prevented the last crisis?
\textit{Journal of Economic Perspectives} 33:107--30

\bibitem[Allen et~al.(2012)Allen, Babus \& Carletti]{allen2012asset}
Allen F, Babus A, Carletti E. 2012.
Asset commonality, debt maturity and systemic risk.
\textit{Journal of Financial Economics} 104:519--534

\bibitem[Allen \& Gale(2000)]{alleng2000}
Allen F, Gale D. 2000.
Financial contagion.
\textit{Journal of Political Economy} 108:1--33

\bibitem[Allen \& Gale(2007)]{allen2007introduction}
Allen F, Gale DM. 2007.
An introduction to financial crises.
\textit{Wharton Financial Institutions Center Working Paper}

\bibitem[Allen et~al.(2006)Allen, Morris \& Shin]{allen2006beauty}
Allen F, Morris S, Shin HS. 2006.
Beauty contests and iterated expectations in asset markets.
\textit{The Review of Financial Studies} 19:719--752

\bibitem[Allouch \& Jalloul(2017)]{allouch2017strategic}
Allouch N, Jalloul M. 2017.
Strategic default in financial networks.
Tech. rep., School of Economics Discussion Papers

\bibitem[Alvarez \& Barlevy(2015)]{alvarezb2015}
Alvarez F, Barlevy G. 2015.
Mandatory disclosure and financial contagion.
\textit{Paper No. w21328, National Bureau of Economic Research}

\bibitem[Amini et~al.(2016)Amini, Cont \& Minca]{amini2016resilience}
Amini H, Cont R, Minca A. 2016.
Resilience to contagion in financial networks.
\textit{Mathematical finance} 26:329--365

\bibitem[Anderson et~al.(2020)Anderson, Erol \&
  Ordo{\~n}ez]{anderson2020interbank}
Anderson H, Erol S, Ordo{\~n}ez G. 2020.
Interbank networks in the shadows of the federal reserve act.
Tech. rep.

\bibitem[Anderson et~al.(2019)Anderson, Paddrik \& Wang]{anderson2019bank}
Anderson H, Paddrik M, Wang JJ. 2019.
Bank networks and systemic risk: Evidence from the national banking acts.
\textit{American Economic Review} 109:3125--61

\bibitem[Atkisson et~al.(2020)Atkisson, G{\'o}rski, Jackson, Ho{\l}yst \&
  D'Souza]{atkisson2020understanding}
Atkisson C, G{\'o}rski PJ, Jackson MO, Ho{\l}yst JA, D'Souza RM. 2020.
Why understanding multiplex social network structuring processes will help us
  better understand the evolution of human behavior.
\textit{Evolutionary Anthropology: Issues, News, and Reviews} 29:102--107

\bibitem[Babus(2016)]{babus2016}
Babus A. 2016.
The formation of financial networks.
\textit{The RAND Journal of Economics} 47:239--272

\bibitem[Babus \& Hu(2017)]{babush2017}
Babus A, Hu TW. 2017.
Endogenous intermediation in over-the-counter markets.
\textit{Journal of Financial Economics} 125:200--215

\bibitem[Banerjee(1992)]{banerjee1992}
Banerjee AV. 1992.
{A simple model of herd behavior}.
\textit{The Quarterly Journal of Economics} :797--817

\bibitem[Bardoscia et~al.(2017)Bardoscia, Battiston, Caccioli \&
  Caldarelli]{bardoscia2017}
Bardoscia M, Battiston S, Caccioli F, Caldarelli G. 2017.
Pathways towards instability in financial networks.
\textit{Nature Communications} 8:14416

\bibitem[Bardoscia et~al.(2018)Bardoscia, Bianconi \& Ferrara]{bardoscia2018}
Bardoscia M, Bianconi G, Ferrara G. 2018.
Multiplex network analysis of the uk otc derivatives market.
\textit{Staff Working Paper: Bank of England}

\bibitem[Barrot \& Sauvagnat(2016)]{barrot2016input}
Barrot JN, Sauvagnat J. 2016.
Input specificity and the propagation of idiosyncratic shocks in production
  networks.
\textit{The Quarterly Journal of Economics} 131:1543--1592

\bibitem[BCBS(2015)]{BCBS2015}
BCBS. 2015.
Making supervisory stress tests more macroprudential: Considering liquidity and
  solvency interactions and systemic risk.
\textit{Basel Committee on Banking Supervision Working Paper 29}

\bibitem[Bebchuk \& Goldstein(2011)]{bebchuk2011self}
Bebchuk LA, Goldstein I. 2011.
Self-fulfilling credit market freezes.
\textit{The Review of Financial Studies} 24:3519--3555

\bibitem[Bech \& Atalay(2010)]{bech2010}
Bech ML, Atalay E. 2010.
The topology of the federal funds market.
\textit{Physica A: Statistical Mechanics and its Applications} 389:5223--5246

\bibitem[Belhaj et~al.(2020)Belhaj, Bourl{\`e}s \&
  Dero{\"\i}an]{belhaj2020prudential}
Belhaj M, Bourl{\`e}s R, Dero{\"\i}an F. 2020.
Prudential regulation in financial networks.
\textit{HAL Id: halshs-02950881
  https://halshs.archives-ouvertes.fr/halshs-02950881}

\bibitem[Bernard et~al.(2017)Bernard, Capponi \& Stiglitz]{bernard2017bail}
Bernard B, Capponi A, Stiglitz JE. 2017.
Bail-ins and bail-outs: Incentives, connectivity, and systemic stability.
Tech. rep., National Bureau of Economic Research

\bibitem[Bikhchandani et~al.(1992)Bikhchandani, Hirshleifer \&
  Welch]{bikhchandanihw1992}
Bikhchandani S, Hirshleifer D, Welch I. 1992.
{A theory of fads, fashion, custom, and cultural change as informational
  cascades}.
\textit{Journal of political Economy} 100

\bibitem[Billio et~al.(2012)Billio, Getmansky, Lo \& Pelizzon]{billioetal2012}
Billio M, Getmansky M, Lo AW, Pelizzon L. 2012.
Econometric measures of connectedness and systemic risk in the finance and
  insurance sectors.
\textit{Journal of Financial Economics} 104(3):535--559

\bibitem[Blasques et~al.(2018)Blasques, Br{\"a}uning \&
  Van~Lelyveld]{blasques2018}
Blasques F, Br{\"a}uning F, Van~Lelyveld I. 2018.
A dynamic network model of the unsecured interbank lending market.
\textit{Journal of Economic Dynamics and Control} 90:310--342

\bibitem[Branch(2002)]{branch2002}
Branch B. 2002.
The costs of bankruptcy: A review.
\textit{International Review of Financial Analysis} 11:1:39--57

\bibitem[Bruche \& Gonzalez-Aguado(2010)]{brucheg2010}
Bruche M, Gonzalez-Aguado C. 2010.
Recovery rates, default probabilities, and the credit cycle.
\textit{Journal of Banking \& Finance} 34:754--764

\bibitem[Brunnermeier(2009)]{brunnermeier2009}
Brunnermeier MK. 2009.
Deciphering the liquidity and credit crunch 2007-2008.
\textit{Journal of Economic perspectives} 23:77--100

\bibitem[Brusco \& Castiglionesi(2007)]{bruscoc2007}
Brusco S, Castiglionesi F. 2007.
Liquidity coinsurance, moral hazard, and financial contagion.
\textit{The Journal of Finance} 62:2275--2302

\bibitem[Brusco \& Jackson(1999)]{bruscoj1999}
Brusco S, Jackson MO. 1999.
The optimal design of a market.
\textit{Journal of Economic Theory} 88:1--39

\bibitem[Burkholz et~al.(2016)Burkholz, Leduc, Garas \&
  Schweitzer]{burkholz2016systemic}
Burkholz R, Leduc MV, Garas A, Schweitzer F. 2016.
Systemic risk in multiplex networks with asymmetric coupling and threshold
  feedback.
\textit{Physica D: Nonlinear Phenomena} 323:64--72

\bibitem[Caballero \& Simsek(2013)]{caballeros2013}
Caballero RJ, Simsek A. 2013.
Fire sales in a model of complexity.
\textit{The Journal of Finance} 68:2549--2587

\bibitem[Cabrales et~al.(2017)Cabrales, Gottardi \&
  Vega-Redondo]{cabralesgv2017}
Cabrales A, Gottardi P, Vega-Redondo F. 2017.
Risk sharing and contagion in networks.
\textit{The Review of Financial Studies} 30:3086--3127

\bibitem[Callaway et~al.(2000)Callaway, Newman, Strogatz \&
  Watts]{callawayetal2000}
Callaway DS, Newman ME, Strogatz SH, Watts DJ. 2000.
Network robustness and fragility: Percolation on random graphs.
\textit{Physical review letters} 85:5468

\bibitem[Calomiris \& Jaremski(2019)]{calomirisj2019}
Calomiris CW, Jaremski M. 2019.
Stealing deposits: Deposit insurance, risk-taking, and the removal of market
  discipline in early 20th-century banks.
\textit{The Journal of Finance} 74:711--754

\bibitem[Capponi \& Chen(2015)]{capponi2015systemic}
Capponi A, Chen PC. 2015.
Systemic risk mitigation in financial networks.
\textit{Journal of Economic Dynamics and Control} 58:152--166

\bibitem[Capponi \& Cheng(2018)]{capponi2018clearinghouse}
Capponi A, Cheng WA. 2018.
Clearinghouse margin requirements.
\textit{Operations Research} 66:1542--1558

\bibitem[Capponi et~al.(2020)Capponi, Corell \& Stiglitz]{capponi2020optimal}
Capponi A, Corell FC, Stiglitz JE. 2020.
Optimal bailouts and the doom loop with a financial network.
Tech. rep., National Bureau of Economic Research

\bibitem[Capponi \& Larsson(2015)]{capponi2015price}
Capponi A, Larsson M. 2015.
Price contagion through balance sheet linkages.
\textit{The Review of Asset Pricing Studies} 5:227--253

\bibitem[Centola(2018)]{centola2018}
Centola D. 2018.
How behavior spreads: The science of complex contagions, vol.~3.
Princeton University Press

\bibitem[Chincarini(2012)]{chincarini2012}
Chincarini LB. 2012.
The crisis of crowding: Quant copycats, ugly models, and the new crash normal.
John Wiley \& Sons

\bibitem[Cifuentes et~al.(2005)Cifuentes, Ferrucci \& Shin]{cifuentesfs2005}
Cifuentes R, Ferrucci G, Shin HS. 2005.
Liquidity risk and contagion.
\textit{Journal of the European Economic Association} 3:556--566

\bibitem[Cohen-Cole et~al.(2015)Cohen-Cole, Patacchini \&
  Zenou]{cohen2015static}
Cohen-Cole E, Patacchini E, Zenou Y. 2015.
Static and dynamic networks in interbank markets.
\textit{Network Science} 3:98--123

\bibitem[Covi et~al.(2018)Covi, Gorpe \& Kok]{covigk2018}
Covi G, Gorpe MZ, Kok C. 2018.
Comap: Mapping contagion in the euro area banking sector.
\textit{mimeo: ECB and IMF}

\bibitem[Craig \& Von~Peter(2014)]{craigv2014}
Craig B, Von~Peter G. 2014.
Interbank tiering and money center banks.
\textit{Journal of Financial Intermediation} 23:322--347

\bibitem[Cs{\'o}ka \& Herings(2018)]{csokah2018}
Cs{\'o}ka P, Herings PJJ. 2018.
Decentralized clearing in financial networks.
\textit{Management Science} 64(10):4681--99

\bibitem[Dam \& Koetter(2012)]{damk2012}
Dam L, Koetter M. 2012.
Bank bailouts and moral hazard: Evidence from germany.
\textit{The Review of Financial Studies} 25:2343--2380

\bibitem[Davydenko et~al.(2012)Davydenko, Strebulaev \& Zhao]{davydenkosz2012}
Davydenko SA, Strebulaev IA, Zhao X. 2012.
A market-based study of the cost of default.
\textit{The Review of Financial Studies} 25:2959--2999

\bibitem[Demange(2016)]{demange2016}
Demange G. 2016.
Contagion in financial networks: a threat index.
\textit{Management Science} 64:955--70

\bibitem[Demsetz(1968)]{demsetz1968cost}
Demsetz H. 1968.
The cost of transacting.
\textit{The quarterly journal of economics} 82:33--53

\bibitem[D'Errico \& Roukny(2019)]{derricor2019}
D'Errico M, Roukny T. 2019.
Compressing over-the-counter markets.
\textit{arXiv preprint arXiv:1705.07155}

\bibitem[Diamond(1991)]{diamond1991monitoring}
Diamond DW. 1991.
Monitoring and reputation: The choice between bank loans and directly placed
  debt.
\textit{Journal of political Economy} 99:689--721

\bibitem[Diamond \& Dybvig(1983)]{diamondd1983}
Diamond DW, Dybvig PH. 1983.
Bank runs, deposit insurance, and liquidity.
\textit{Journal of political economy} 91:401--419

\bibitem[Diamond \& Rajan(2011)]{diamondr2011}
Diamond DW, Rajan RG. 2011.
Fear of fire sales, illiquidity seeking, and credit freezes.
\textit{The Quarterly Journal of Economics} 126:557--591

\bibitem[Diebold \& Y{\i}lmaz(2014)]{dieboldy2014}
Diebold FX, Y{\i}lmaz K. 2014.
On the network topology of variance decompositions: Measuring the connectedness
  of financial firms.
\textit{Journal of Econometrics} 182:119--134

\bibitem[Duarte \& Eisenbach(2018)]{duarte2018fire}
Duarte F, Eisenbach TM. 2018.
Fire-sale spillovers and systemic risk.
\textit{FRB of New York Staff Report}

\bibitem[Duarte \& Jones(2017)]{duartej2017}
Duarte F, Jones C. 2017.
Empirical network contagion for us financial institutions.
\textit{Federal Reserve Bank of New York Staff Reports}

\bibitem[Duffie(2019)]{duffie2019prone}
Duffie D. 2019.
Prone to fail: The pre-crisis financial system.
\textit{Journal of Economic Perspectives} 33:81--106

\bibitem[Duffie et~al.(2009)Duffie, Eckner, Horel \& Saita]{duffieeetal2009}
Duffie D, Eckner A, Horel G, Saita L. 2009.
Frailty correlated default.
\textit{The Journal of Finance} 64:2089--2123

\bibitem[Duffie \& Wang(2016)]{duffiew2016}
Duffie D, Wang C. 2016.
Efficient contracting in network financial markets.
\textit{working paper: Stanford}

\bibitem[Duffie \& Zhu(2011)]{duffiez2011}
Duffie D, Zhu H. 2011.
Does a central clearing counterparty reduce counterparty risk?
\textit{The Review of Asset Pricing Studies} 1:74--95

\bibitem[Eisenberg \& Noe(2001)]{eisenbergn2001}
Eisenberg L, Noe TH. 2001.
Systemic risk in financial systems.
\textit{Management Science} 47:236--249

\bibitem[Elliott et~al.(2018)Elliott, Georg \& Hazell]{elliottgh2018}
Elliott M, Georg CP, Hazell J. 2018.
Systemic risk-shifting in financial networks.
\textit{mimeo: Cambridge University}

\bibitem[Elliott et~al.(2014)Elliott, Golub \& Jackson]{elliottgj2014}
Elliott M, Golub B, Jackson MO. 2014.
Financial networks and contagion.
\textit{American Economic Review} 104(10):3115--3153

\bibitem[Engle \& Ruan(2019)]{engle2019measuring}
Engle RF, Ruan T. 2019.
Measuring the probability of a financial crisis.
\textit{Proceedings of the National Academy of Sciences} 116:18341--18346

\bibitem[Erol(2019)]{erol2019}
Erol S. 2019.
Network hazard and bailouts.
\textit{Available at SSRN 3034406}

\bibitem[Erol \& Vohra(2018)]{erolv2018}
Erol S, Vohra R. 2018.
Network formation and systemic risk.
\textit{mimeo: UPenn}

\bibitem[Etessami et~al.(2019)Etessami, Papadimitriou, Rubinstein \&
  Yannakakis]{etessami2019tarski}
Etessami K, Papadimitriou C, Rubinstein A, Yannakakis M. 2019.
Tarski's theorem, supermodular games, and the complexity of equilibria.
\textit{arXiv preprint arXiv:1909.03210}

\bibitem[Farboodi(2017)]{farboodi2017}
Farboodi M. 2017.
Intermediation and voluntary exposure to counterparty risk.
\textit{mimeo:}

\bibitem[Farmer et~al.(2020)Farmer, Kleinnijenhuis, Nahai-Williamson \&
  Wetzer]{farmer2020foundations}
Farmer JD, Kleinnijenhuis AM, Nahai-Williamson P, Wetzer T. 2020.
Foundations of system-wide financial stress testing with heterogeneous
  institutions.
\textit{Bank of England Working Paper}

\bibitem[Ferrara et~al.(2017)Ferrara, Langfield, Liu \& Ota]{ferrara2017}
Ferrara G, Langfield S, Liu Z, Ota T. 2017.
Systemic illiquidity in the interbank network.
\textit{Staff Working Paper: Bank of England}

\bibitem[Fleming \& Sarkar(2014)]{flemings2014}
Fleming MJ, Sarkar A. 2014.
The failure resolution of lehman brothers.
\textit{Economic Policy Review, Federal Reserve Bank of New York} 20

\bibitem[Fostel \& Geanakoplos(2008)]{fostel2008leverage}
Fostel A, Geanakoplos J. 2008.
Leverage cycles and the anxious economy.
\textit{American Economic Review} 98:1211--44

\bibitem[Fostel \& Geanakoplos(2014)]{fostel2014endogenous}
Fostel A, Geanakoplos J. 2014.
Endogenous collateral constraints and the leverage cycle.
\textit{Annu. Rev. Econ.} 6:771--799

\bibitem[Fricke \& Wilke(2020)]{fricke2020connected}
Fricke D, Wilke H. 2020.
Connected funds.
\textit{mimeo}

\bibitem[Gai et~al.(2011)Gai, Haldane \& Kapadia]{gai-haldane-kapadia}
Gai P, Haldane A, Kapadia S. 2011.
Complexity, concentration and contagion.
\textit{Journal of Monetary Economics} 58:5:453--470

\bibitem[Gai \& Kapadia(2010)]{gaik2010}
Gai P, Kapadia S. 2010.
Contagion in financial networks.
\textit{Proceedings of the Royal Society A} 466:2401--2423

\bibitem[Gale \& Kariv(2007)]{galek2007}
Gale DM, Kariv S. 2007.
Financial networks.
\textit{American Economic Review} 97:99--103

\bibitem[Galeotti \& Ghiglino(2019)]{galeottig2019}
Galeotti A, Ghiglino C. 2019.
Cross-ownership and portfolio choice.
\textit{mimeo}

\bibitem[Garas(2016)]{garas2016interconnected}
Garas A. 2016.
Interconnected networks.
Springer

\bibitem[Gehrig(1993)]{gehrig1993}
Gehrig T. 1993.
Intermediation in search markets.
\textit{Journal of Economics \& Management Strategy} 2:97--120

\bibitem[Glasserman \& Young(2015)]{glassermany2015}
Glasserman P, Young HP. 2015.
How likely is contagion in financial networks?
\textit{Journal of Banking \& Finance} 50:383--399

\bibitem[Glasserman \& Young(2016)]{glassermany2016}
Glasserman P, Young HP. 2016.
Contagion in financial networks.
\textit{Journal of Economic Literature} 54:3:779--831

\bibitem[Glode \& Opp(2016)]{glode2016asymmetric}
Glode V, Opp C. 2016.
Asymmetric information and intermediation chains.
\textit{American Economic Review} 106:2699--2721

\bibitem[Glode et~al.(2019)Glode, Opp \& Zhang]{glode2019efficiency}
Glode V, Opp CC, Zhang X. 2019.
On the efficiency of long intermediation chains.
\textit{Journal of Financial Intermediation} 38:11--18

\bibitem[Godlewski \& Sanditov(2018)]{godlewski2018financial}
Godlewski CJ, Sanditov B. 2018.
Financial institutions network and the certification value of bank loans.
\textit{Financial management} 47:253--283

\bibitem[Godlewski et~al.(2012)Godlewski, Sanditov \&
  Burger-Helmchen]{godlewski2012bank}
Godlewski CJ, Sanditov B, Burger-Helmchen T. 2012.
Bank lending networks, experience, reputation, and borrowing costs: empirical
  evidence from the french syndicated lending market.
\textit{Journal of Business Finance \& Accounting} 39:113--140

\bibitem[Gofman(2017)]{gofman2017}
Gofman M. 2017.
Efficiency and stability of a financial architecture with
  too-interconnected-to-fail institutions.
\textit{Journal of Financial Economics} 124:113--146

\bibitem[Greenwood et~al.(2015)Greenwood, Landier \&
  Thesmar]{greenwood2015vulnerable}
Greenwood R, Landier A, Thesmar D. 2015.
Vulnerable banks.
\textit{Journal of Financial Economics} 115:471--485

\bibitem[Gualdi et~al.(2016)Gualdi, Cimini, Primicerio, Di~Clemente \&
  Challet]{gualdi2016}
Gualdi S, Cimini G, Primicerio K, Di~Clemente R, Challet D. 2016.
Statistically validated network of portfolio overlaps and systemic risk.
\textit{Scientific reports} 6:39467

\bibitem[Haldane(2009)]{haldane2009}
Haldane AG. 2009.
Rethinking the financial network.
\textit{Speech delivered at the Financial Student Association in Amsterdam, The
  Netherlands, April 2009}

\bibitem[Hastings et~al.(2020)Hastings, Hemenway~Falk \&
  Tsoukalas]{hastings2020privacy}
Hastings M, Hemenway~Falk B, Tsoukalas G. 2020.
Privacy-preserving network analytics.
\textit{Available at SSRN: http://dx.doi.org/10.2139/ssrn.3680000}

\bibitem[Hauton \& H{\'e}am(2016)]{hauton2016}
Hauton G, H{\'e}am JC. 2016.
How to measure interconnectedness between banks, insurers and financial
  conglomerates.
\textit{Statistics \& Risk Modeling} 33:95--116

\bibitem[Heipertz et~al.(2019)Heipertz, Ouazad \&
  Ranci{\`e}re]{heipertz2019transmission}
Heipertz J, Ouazad A, Ranci{\`e}re R. 2019.
The transmission of shocks in endogenous financial networks: A structural
  approach.
Tech. rep., National Bureau of Economic Research

\bibitem[Herskovic(2018)]{herskovic2018networks}
Herskovic B. 2018.
Networks in production: Asset pricing implications.
\textit{The Journal of Finance} 73:1785--1818

\bibitem[Hirshleifer \& Teoh(2009)]{hirshleifert2009}
Hirshleifer D, Teoh SH. 2009.
Systemic risk, coordination failures, and preparedness externalities:
  Applications to tax and accounting policy.
\textit{Journal of Financial Economic Policy} 1:128--142

\bibitem[Holmstrom \& Tirole(1997)]{holmstrom1997financial}
Holmstrom B, Tirole J. 1997.
Financial intermediation, loanable funds, and the real sector.
\textit{the Quarterly Journal of economics} 112:663--691

\bibitem[Ibragimov et~al.(2011)Ibragimov, Jaffee \&
  Walden]{ibragimov2011diversification}
Ibragimov R, Jaffee D, Walden J. 2011.
Diversification disasters.
\textit{Journal of Financial Economics} 99:333--348

\bibitem[Infante \& Vardoulakis(2018)]{infante2018collateral}
Infante S, Vardoulakis A. 2018.
Collateral runs.
\textit{Available at SSRN 3099637}

\bibitem[Jackson(2008)]{jackson2008}
Jackson MO. 2008.
Social and economic networks.
Princeton: Princeton University Press

\bibitem[Jackson(2019)]{jackson2019}
Jackson MO. 2019.
The human network: How your social position determines your power, beliefs and
  behaviors.
Pantheon Books: New York

\bibitem[Jackson \& Nei(2015)]{jacksonn2015}
Jackson MO, Nei S. 2015.
Networks of military alliances, wars, and international trade.
\textit{Proceedings of the National Academy of Sciences} 112(50):15277--15284

\bibitem[Jackson \& Pernoud(2019)]{jacksonp2019}
Jackson MO, Pernoud A. 2019.
Distorted investment incentives, regulation, and equilibrium multiplicity in a
  model of financial networks.
\textit{SSRN: https://dx.doi.org/10.2139/ssrn.3311839}

\bibitem[Jackson \& Pernoud(2020)]{jacksonp2020}
Jackson MO, Pernoud A. 2020.
Credit freezes, equilibrium multiplicity, and optimal bailouts in financial
  networks.
\textit{SSRN: https://ssrn.com/abstract=3735251}

\bibitem[Jackson \& Storms(2017)]{jacksons2017}
Jackson MO, Storms EC. 2017.
Behavioral communities and the atomic structure of networks.
\textit{Arxiv: https://arxiv.org/abs/1710.04656}

\bibitem[Jackson \& Wolinsky(1996)]{jacksonw1996}
Jackson MO, Wolinsky A. 1996.
A strategic model of social and economic networks.
\textit{Journal of Economic Theory} 71:44--74

\bibitem[James(1991)]{james1991}
James C. 1991.
The losses realized in bank failures.
\textit{The Journal of Finance} 46:1223--1242

\bibitem[Jensen \& Meckling(1976)]{jensenm1976}
Jensen MC, Meckling WH. 1976.
Theory of the firm: Managerial behavior, agency costs and ownership structure.
\textit{Journal of financial economics} 3:305--360

\bibitem[Kanik(2019)]{kanik2018}
Kanik Z. 2019.
From lombard street to wall street: Systemic risk, rescues, and stability in
  financial networks.
\textit{mimeo: MIT}

\bibitem[Karamysheva \& Seregina(2020)]{karamyshevas2020}
Karamysheva M, Seregina E. 2020.
Prudential policies and systemic risk: the role of interconnections.
\textit{mimeo: HSE Moscow}

\bibitem[Keynes(1936)]{keynes1936}
Keynes JM. 1936.
The general theory of employment, interest, and money.
London: Macmillan

\bibitem[King \& Wadhwani(1990)]{king1990transmission}
King MA, Wadhwani S. 1990.
Transmission of volatility between stock markets.
\textit{The Review of Financial Studies} 3:5--33

\bibitem[Kivela et~al.(2014)Kivela, Arenas, Gleeson, Moreno \&
  Porter]{kivelaetal2014}
Kivela M, Arenas A, Gleeson JP, Moreno Y, Porter MA. 2014.
Multilayer networks.
\textit{arXiv:1309.7233v4 [physics.soc-ph]}

\bibitem[Kiyotaki \& Moore(1997)]{kiyotakim1997}
Kiyotaki N, Moore J. 1997.
Credit cycles.
\textit{Journal of Political Economy} 105(2):211 -- 248

\bibitem[Klasing \& Milionis(2014)]{klasingm2014}
Klasing MJ, Milionis P. 2014.
Quantifying the evolution of world trade, 1870--1949.
\textit{Journal of International Economics} 92:185--197

\bibitem[Krishnamurthy(2010)]{krishnamurthy2010amplification}
Krishnamurthy A. 2010.
Amplification mechanisms in liquidity crises.
\textit{American Economic Journal: Macroeconomics} 2:1--30

\bibitem[Leitner(2005)]{leitner2005}
Leitner Y. 2005.
Financial networks: Contagion, commitment, and private sector bailouts.
\textit{Journal of Finance} 60(6):2925--2953

\bibitem[Lucas(2019)]{lucas2019measuring}
Lucas D. 2019.
Measuring the cost of bailouts.
\textit{Annual Review of Financial Economics} 11:85--108

\bibitem[Lucas(2013)]{lucas2013}
Lucas Jr. RE. 2013.
Glass-steagall: A requiem.
\textit{American Economic Review: Papers and Proceedings} 103:3:43--47

\bibitem[Lund \& H{\"a}rle(2017)]{lund2017global}
Lund S, H{\"a}rle P. 2017.
Global finance resets.
\textit{Finance and development (IMF Publication)} 54(4):42--44

\bibitem[Malherbe(2014)]{malherbe2014self}
Malherbe F. 2014.
Self-fulfilling liquidity dry-ups.
\textit{The Journal of Finance} 69:947--970

\bibitem[Martinezy et~al.(2020)Martinezy, Peirisz \&
  Tsomocos]{martinezetal2020}
Martinezy JF, Peirisz M, Tsomocos D. 2020.
Macroprudential policy analysis in an estimated dsge model with a heterogeneous
  banking system: an application to chile.
\textit{mimeo: HSE Moscow}

\bibitem[Morris \& Shin(1998)]{morriss1998}
Morris S, Shin HS. 1998.
Unique equilibrium in a model of self-fulfilling currency attacks.
\textit{American Economic Review} :587--597

\bibitem[Morris \& Shin(2002)]{morris2002social}
Morris S, Shin HS. 2002.
Social value of public information.
\textit{american economic review} 92:1521--1534

\bibitem[Nanumyan et~al.(2015)Nanumyan, Garas \&
  Schweitzer]{nanumyan2015network}
Nanumyan V, Garas A, Schweitzer F. 2015.
The network of counterparty risk: Analysing correlations in otc derivatives.
\textit{PloS one} 10:e0136638

\bibitem[O'hara(1997)]{o1997market}
O'hara M. 1997.
Market microstructure theory.
Wiley

\bibitem[Rajan(1992)]{rajan1992insiders}
Rajan RG. 1992.
Insiders and outsiders: The choice between informed and arm's-length debt.
\textit{The Journal of finance} 47:1367--1400

\bibitem[Ram{\'\i}rez(2019)]{ramirez2019}
Ram{\'\i}rez C. 2019.
Regulating financial networks under uncertainty.
\textit{FEDS Working Paper}

\bibitem[Ram{\'\i}rez(2017)]{ramirez2017}
Ram{\'\i}rez CA. 2017.
Firm networks and asset returns

\bibitem[Reinhart \& Rogoff(2009)]{reinhartr2009}
Reinhart C, Rogoff K. 2009.
This time is different.
Princeton University Press: Princeton

\bibitem[Rochet \& Tirole(1996)]{rochett1996}
Rochet JC, Tirole J. 1996.
Interbank lending and systemic risk.
\textit{Journal of Money, Credit and Banking} 28:4:733--762

\bibitem[Rogers \& Veraart(2013)]{rogers2013failure}
Rogers LC, Veraart LA. 2013.
Failure and rescue in an interbank network.
\textit{Management Science} 59:882--898

\bibitem[Roukny et~al.(2018)Roukny, Battiston \& Stiglitz]{rouknybs2018}
Roukny T, Battiston S, Stiglitz JE. 2018.
Interconnectedness as a source of uncertainty in systemic risk.
\textit{Journal of Financial Stability} 35:93--106

\bibitem[Scharfstein \& Stein(1990)]{scharfsteins1990}
Scharfstein DS, Stein JC. 1990.
Herd behavior and investment.
\textit{The American Economic Review} :465--479

\bibitem[Shell(1989)]{shell1989sunspot}
Shell K. 1989.
Sunspot equilibrium. In \textit{General Equilibrium}. Springer,  274--280

\bibitem[Shiller(2015)]{shiller2015}
Shiller RJ. 2015.
Irrational exuberance: Revised and expanded third edition.
Princeton university press

\bibitem[Shu(2019)]{shu2019}
Shu C. 2019.
Endogenous risk-exposure and systemic instability.
\textit{USC-INET Research Paper}

\bibitem[Siebenbrunner(2020)]{siebenbrunner2020quantifying}
Siebenbrunner C. 2020.
Quantifying the importance of different contagion channels as sources of
  systemic risk.
\textit{Journal of Economic Interaction and Coordination} :1--29

\bibitem[Silber(1983)]{silber1983process}
Silber WL. 1983.
The process of financial innovation.
\textit{The American Economic Review} 73:89--95

\bibitem[Soram\"aki et~al.(2007)Soram\"aki, Bech, Arnold, Glass \&
  Beyeler]{soramakietal2007}
Soram\"aki K, Bech ML, Arnold J, Glass RJ, Beyeler WE. 2007.
The topology of interbank payment flows.
\textit{Physica A} 379:317--333

\bibitem[Spulber(1996)]{spulber1996market}
Spulber DF. 1996.
Market microstructure and intermediation.
\textit{Journal of Economic perspectives} 10:135--152

\bibitem[Stellian et~al.(2020)Stellian, Penagos \&
  Danna-Buitrago]{stellian2020firms}
Stellian R, Penagos GI, Danna-Buitrago JP. 2020.
Firms in financial distress: evidence from inter-firm payment networks with
  volatility driven by ‘animal spirits’.
\textit{Journal of Economic Interaction and Coordination} :1--43

\bibitem[Summer(2013)]{summer2013}
Summer M. 2013.
Financial contagion and network analysis.
\textit{Annual Review of Financial Economics} 5:277--297

\bibitem[Tarullo(2019)]{tarullo2019financial}
Tarullo DK. 2019.
Financial regulation: Still unsettled a decade after the crisis.
\textit{Journal of Economic Perspectives} 33:61--80

\bibitem[Teteryatnikova(2014)]{teteryatnikova2014systemic}
Teteryatnikova M. 2014.
Systemic risk in banking networks: Advantages of “tiered” banking systems.
\textit{Journal of Economic Dynamics and Control} 47:186--210

\bibitem[Upper \& Worms(2004)]{upperw2004}
Upper C, Worms A. 2004.
Estimating bilateral exposures in the german interbank market: Is there a
  danger of contagion?
\textit{European Economic Review} 48:827--849

\bibitem[Wagner(2010)]{wagner2010}
Wagner W. 2010.
Diversification at financial institutions and systemic crises.
\textit{Journal of Financial Intermediation} 19:373--386

\bibitem[Wang(2017)]{wang2017}
Wang C. 2017.
Core-periphery trading networks.
\textit{Dissertation, Stanford University}

\bibitem[Wang et~al.(2020)Wang, Capponi \& Zhang]{wang2020theory}
Wang JJ, Capponi A, Zhang H. 2020.
A theory of collateral requirements for central counterparties

\bibitem[Yellen(2013)]{yellen2013interconnectedness}
Yellen J. 2013.
Interconnectedness and systemic risk: Lessons from the financial crisis and
  policy implications: a speech at the american economic association/american
  finance association joint luncheon, san diego, california, january 4, 2013.
Tech. rep., Board of Governors of the Federal Reserve System (US)

\end{thebibliography}

\setcounter{page}{0}\thispagestyle{empty}
\newpage

{\bf  \noindent Supplemental Online Appendix to
\\    Systemic Risk in Financial Networks:  A Survey \\  Jackson and Pernoud }

\bigskip

\section{Some Background on Financial Interconnectedness}

Before getting into the review of the literature on systemic risk and financial networks,
it is useful to provide some empirical background and context, which is the purpose of this section.
 Anyone who reads  Janet Yellen's  \citeyearpar{yellen2013interconnectedness}
speech,  or the book by \cite*{reinhartr2009}, will realize that systemic risk is not a new phenomenon.
Nonetheless, the extent of the network across borders has increased, and this means that a crisis in one
country can quickly become an international crisis; and the 2008 financial crisis illustrated
the global nature of our financial network quite clearly.   In addition, the size and centrality of some of
the largest players is reaching an unprecedented scale, and at the same time the network is becoming more
complex and harder to track, as it
involves more diversity and specialization than ever.
We detail these facts below.   We also mention two other important background facts that are
less about how the financial system has evolved, but instead provide new information
about potential dangers and costs in the system.   The first concerns
the high correlation in portfolios that linked financial institutions hold -- which leads to increased dangers of cascades
due to correlated times at which financial institutions are vulnerable; and the second concerns the size of bankruptcy costs --
which are one important part of the economic damage that cascades in a financial crisis.

\subsection{Globalization and Financial Interdependencies}\label{global}

World trade grew from just under 20 percent of world GDP at the end of the Second World War to over 60 percent by 2015.\footnote{Imports plus exports over GDP.  Detailed data can be found for 1870-1949: \cite{klasingm2014}; 1950-1959: Penn World Trade Tables Version 8.1; 1960-2015: World Bank World Development Indicators.
}
It is not a coincidence that the world poverty rate fell from over 40 percent in the early 1980s to below 10 percent in 2020.\footnote{World Bank Poverty Report.}
This trend was matched by a {tenfold} decrease in the incidence of wars, by multiple measures:  effectively, trading partners almost never go to war with each other.\footnote{See \cite*{jacksonn2015} for more empirical background and analysis of the relationship between increased trade and decreased interstate armed conflict.}
These enormous benefits mean that there are strong reasons that we should welcome the very large network that has emerged.
Nonetheless, such growth in interconnectedness comes with risks of systemic disruptions - both via the increasingly complex supply chains and
the financial networks that support the whole system.

Indeed, the accompanying growth in cross border finances is impressive.
For example,
17 percent of equities and 18 percent of bonds around the world were held by foreigners in 2000, and
that rose to 27 percent of equities and 31 percent of bonds by 2016.
This matches up with
the investments (debt, equity, FDI, lend/other) around the world
that come from foreign sources; which
at the more than 132 trillion dollars in 2016 (\cite{lund2017global}) --  compared to a total level of world investments of just over 300 trillion dollars --
is well more than a third of all finances.

\subsection{Consolidation}\label{consolidation}

The financial/banking sector has grown enormously, but has also consolidated,  with far fewer banks and those being much larger than they used to be.
In 1980 there were 14 thousand commercial banks in the US according to the FDIC\footnote{See https://www5.fdic.gov/hsob/HSOBRpt.asp }, with total assets of 2 trillion dollars.  In 2018 there were 4.7K with 16.5 trillion dollars in assets.
So the number of banks has dropped to a third of what it was, and at the same time banks are managing more than eight times as much in terms of total assets.\footnote{Part of this change is due to changes in regulations, such as undoing the separation of investment, commercial banking, and insurance, that had been required under Glass Steagall.  However, this trend is also seen outside of the US, reflecting large economies of scope and scale in the banking sector.}
This consolidation has continued to grow even after the 2008 financial crisis.
For example,
in 1990 the five largest banks in the US held 10 percent of total financial assets, in 2007 they held 35 percent, and in 2015 45 percent.
The ten largest banks in the world controlled 26 trillion dollars in 2016.  To put that in perspective, the US and Chinese combined GDP in 2016 was 29 trillion dollars, and the world GDP was 75 trillion dollars.

\subsection{The Spectrum and Role of Financial Intermediaries}

It may seem paradoxical that the growth in the size of the banking sector and consolidation in the number of banks have been accompanied by a proliferation in the
number of different types of financial intermediaries and increasing
specialization in their roles.\footnote{A general role of
intermediaries arises from providing liquidity in markets where there can be temporary imbalances betwen buyers and sellers,
or where centralized trade can lower search costs.  For
some background, see \cite{demsetz1968cost,gehrig1993, spulber1996market,o1997market,bruscoj1999}. }

To fix ideas consider the following illustration.
A century ago a mortgage was typically issued by a bank and often it was the sole intermediary between that borrower and the bank's depositors who were the effective lenders.
The bank served a number of roles.
On one end, it took in deposits from people who had different and random times at which they needed their money back, and valued some flexibility in
their ability to withdraw funding.   By pooling the money from many different depositors and taking advantage of laws of large numbers, it could largely predict when
it would need to pay parts of the funding back to its depositors, and hence had reliable streams that it could lend out for fixed durations.\footnote{For basic models of this, see for instance, \cite{diamondd1983,alleng2000}.}
This allowed it to lend money out in a portfolio of mortgages of different maturities and risks.
On this other end, the bank helped the borrower select an appropriate mortgage, and screened the borrower to ensure that he had good credit and that the property and situation was properly evaluated in terms of the risks involved.  It also monitored the loan and collected the payments over time.
There were significant economies of scale not only in pooling the deposits, but also in having an intermediary with the expertise of evaluating prospective borrowers, marketing the mortgages, collecting payments, and balancing the portfolio to match with the depositors' needs.

Over time the chains of parties involved in this financial intermediation has grown as the multiple roles of the bank have been separated.  A mortgage may now be issued through a broker who provides the sales and marketing expertise.   The brokers work with a multitude of firms who do the actual issuing of the mortgages, including banks as well as other companies.  They specialize in documenting the circumstances of the borrower and property involved and then often resell the mortgages.   Many mortgages are purchased and held en masse by entities that collect the payments and then resell
those streams of payments in different tranches (packages of mortgages grouped by risks and maturities) in the form of mortgage-backed securities.   Those securities might be bought by banks and other investment companies who then package them together in portfolios either to pay interest to their depositors or offer them as part of investment funds to private investors.  Along the way, various parties in this chain insure and hedge their risks via a variety of derivatives and insurance contracts that are often sold by other firms that are separate from all the others in the chains.

In this example, the many roles that were filled historically by a single bank have been separated: the broker provides sales expertise, the mortgage company does due diligence on the borrower and property, the next buyer and repackager of the mortgages pools risks and takes advantage of economies of scale to eliminate idiosyncratic risks and provide a more predictable return on a security, the purchasers of those securities provide returns and flexibility to large numbers of depositors who want random access to their funds as well as some return on their investment.\footnote{As we saw in the 2008 financial crisis, there can be failures to properly fulfill their roles by all parts of these chains (e.g., see the discussion in Chapter 4 of \cite{jackson2019}).
There were brokers paid by commission per loan encouraging buyers to take on loans that were inappropriate for them.  Some of the largest firms issuing mortgages did not screen the loans properly and were fraudulent in what they told buyers about what they resold (e.g., Countrywide Mortgage).   Many mortgages were sold to large mortgage warehouses such as Fannie Mae and Freddie Mac (FNMA and FHLMC) who ended up improperly valuing the trillions of dollars worth of mortgages that they were buying.   A variety of asset-backed and mortgage-backed securities were being insured by AIG, which ended up not even having the capital
to make margin payments on some of the contracts it sold.   The securities were bought in large amounts by various banks and investment banks who over-valued them (e.g., Bears Stearns and Lehman Brothers), leading to excess risk and eventual defaults to their investors and depositors.}
In some cases, several of these services are provided by different parts of a single bank holding company, but some are no longer part of the banking sector and have become part of a shadow banking sector.

This is just one example, and there are many other situations that involve multiple intermediaries
including insurance, venture capital, corporate lending, and many other forms of
investments.\footnote{See \cite*{glode2016asymmetric, glode2019efficiency} for models of OTC markets with
asymmetric information, in which longer intermediation chains can sometimes improve trade efficiency.}
In sum, there has been a proliferation of different types of financial intermediaries who specialize in different parts of the chain between the ultimate borrowers and lenders; at the same time as the substantial economies of scale in many of the particular roles have led to consolidation and concentration within each role.

This proliferation leads to a challenge in evaluating systemic risk: the networks involve organizations that fall into different regulatory jurisdictions, and some important players are not
actively monitored at all.

\subsection{Core-Periphery Structures}\label{core}

One microcosm of the different roles of financial intermediaries discussed
above is reflected in the structure of the banking sector itself,
which is often crudely divided into two parts: a core of very large
national/international banks and a periphery of smaller (but often still
large) regional banks.
The core banks are highly interconnected with each other, whereas the rest
of the network is usually sparse, with regional banks each only interacting
with a few of the core banks.
Empirical studies of financial networks, and especially interbank lending,
have highlighted this core-periphery structure.  For instance,
\cite*{soramakietal2007} detail a completely connected core of 25 banks,
including all of the largest ones, that borrow from and lend to each other;
with large exposures between them.
Other studies the document such structures include \cite*{bech2010} for the US,
\cite*{upperw2004} and \cite*{craigv2014} for Germany,
and \cite*{blasques2018} for the Netherlands.
As alluded to above, there are a variety of reasons to expect a core-periphery structure as there are advantages to having a concentration in the core of intermediaries, which
can then use large economies of scale to better manage their inventory and match buyers with sellers (e.g., see \cite*{craigv2014,babush2017,farboodi2017,wang2017}), while the periphery can specialize in local expertise in making loans.\footnote{\cite{wang2017} provides extensive references to studies documenting the presence of core-periphery structures in wide varieties of over-the-counter markets.}

In terms of understanding systemic risk, the dense connections within a core-periphery structure also have contagion consequences.  For instance, \cite*{elliottgj2014} show how the core
can lead to much more extensive default cascades for wider ranges of transactions than more balanced networks.
This is exacerbated by the fact that core organizations often have very similar businesses and thus very correlated investments,
which leads them to be vulnerable at the same time, as we discuss next.

\subsection{Correlated Investments}

The financial crisis of 2008 was an obvious situation in which many financial institutions were heavily exposed to the same mortgage and subprime mortgage markets,
and had extensive exposures to each other at the same time.
Since then, several studies have examined this sort of correlation explicitly, and it continues.\footnote{See \cite*{ibragimov2011diversification} for some analysis of how diversification and risk depend on
the distributions of available returns in various asset classes.}
For instance, \cite*{elliottgh2018} find that German banks are more likely to lend to banks with portfolios similar to their own: going from the 25th to 75th percentile of similarity in portfolios between two banks
increases their lending to each other by 31 percent.  They also find an effect on the extensive margin in
terms of the probability that they lend to each other at all.\footnote{For other discussions
and numbers see, for example, \cite*{duffieeetal2009,dieboldy2014,nanumyan2015network} and \cite*{gualdi2016}. }

This sort of correlation occurs for many reasons.  Four primary ones are as follows.

First, competition between institutions can lead them to choose similar investments.   This was something that contributed to the
savings and loan institutions' extensive exposure to fixed-rate mortgages and later to junk-bonds and
the S\&L crisis of the 1980s and 90s.    This happened since savings and loans that took riskier positions
could offer higher interest rates on their checking and savings accounts (and given that much of their pre-existing
fixed-rate mortgage portfolios were paying below-market returns, they had to find very high return investments to match the high market interest rates that were
prevalent at the time).  Since many of these were insured accounts, depositors had incentives to shop for the highest interest rate.   This means that in order to attract and keep depositors, savings and loans had to compete to offer the highest interest rates.
Since the place they could earn the higher expected rates of return that would enable them to offer higher interest rates was in riskier investments, and in the junk-bond market in particular, this drove them to take increasingly risky positions.
This incentive is not unique to the savings and loan crisis, but is more commonly at work in the banking sector.

Second, there can be regulations restricting the sorts of investments that banks and
other financial institutions can make with some of their capital, or other
requirements on which sorts of assets they can use to satisfy reserve
requirements or can use for short term lending and repos.   If banks need to
hold some percentage of their portfolio in bonds issued by some countries
(e.g., European countries) then it is not surprising that they all have
nontrivial amounts of investment in bonds that offer higher returns (e.g., Greek
debt in 2010, just before their debt crisis).\footnote{This can interact with miscategorization by ratings agencies. If ratings agencies mistakenly list some assets as being of a higher class than they should be, then that may qualify those assets to be used for ``risk-free'' purposes, even though they are risky and are thus offering higher expected returns.}  This can correlate their portfolios and make them susceptible to the same shocks.

Third, as we discuss more extensively below, banks have very strong incentives to deliberately choose portfolios that are correlated with those of their counterparties.  This follows since
 they benefit most from being solvent when they earn the greatest returns from
their counterparties and insolvent when their counterparties are insolvent.

Fourth, regardless of the short-run correlations in bank portfolios, there can also be large economy-wide shocks, such as the lost production and employment due to covid-19, which affect the portfolios of almost all financial institutions in the same direction at the same time.

\subsection{Bankruptcy Costs}\label{bankruptcycosts}

The externalities between financial organizations matter not only because of the basic investment distortions that result,
but also because of the substantial frictions and costs of bankruptcy that are present in financial networks.
If a large counterparty of some financial organization defaults, that can result in large
losses for the organization, and ultimately cause it to default  as well.

These costs are of fundamental importance, since otherwise defaults or
changes in values in the financial system only determine which organizations
have which values, but not the overall total of those values.
Bankruptcy costs due to insolvency are important in leading to extra drops in values, cascades of those losses, and overall depression of values.

As an example of the scale of real economic costs when there are defaults, let us have a look at the Lehman Brothers default.  In that bankruptcy there were initially 1.2 trillion dollars of claims made against Lehman Brothers.
Of these, the courts ultimately allowed only 362 billion dollars of claims, and then those creditors only received 28 percent of that reduced number.\footnote{See \cite*{flemings2014}.}
This was an extreme case,
but there are substantial frictions, delays, and inefficiencies that result from bankruptcy, especially in troubled times.
These result from fire sales, early termination of contracts, the complexity of contracts that need to be unwound, lengthy negotiations, legal costs,
among others.

Estimates of bankruptcy recovery rates are in the 56-57 percent range -- so that more than 40 percent of the value of an organization going into bankruptcy is lost in the process.\footnote{ See \cite*{branch2002,acharyabs2007}, as well as
 \cite*{davydenkosz2012,james1991}.}
Of the amounts that are lost about   4/11 is attributable to legal costs and the other 7/11 to a drop in asset value (some from liquidation).
Moreover, recovery rates are another 15 to 22 percent lower in distressed
times,\footnote{See \cite*{acharyabs2007,brucheg2010}.}
which would typically apply during a large financial crisis.

\newpage
\section{An Executive Summary}\label{exec}

\begin{itemize}
\item Financial intermediation is vital to any economy:
\begin{itemize}
\item It takes advantage of economies of scope and scale in pooling idiosyncracies on both borrowing and lending sides, provides trading opportunities when long and short sides are unbalanced, provides screening and monitoring of investments, and matches funds to investments.
\item However, it also involves substantial externalities as financial intermediaries are interconnected and interact through substantially incomplete markets,
and face substantial bankruptcy costs and inefficiencies due to lost
investments during crises; which cause real damage to the economy,
especially when they cascade.
\item  We discuss the various ways financial interconnections generate systemic risk, the inefficiencies that ensue, as well as when and how oversight, regulation, and interventions are useful.
\end{itemize}

\item Financial intermediaries and other financial institutions are interconnected in several ways:
\begin{itemize}
\item via a variety of explicit contracts that directly expose them to each other;
\item via similar investments and correlation in the values of their portfolios; making them vulnerable
to the same forces  at the same time;
\item via investors' inferences about the health of one institution from the values of others.
\end{itemize}

\item Correspondingly, systemic risk comes from several forms of contagion:
\begin{itemize}
\item  via cascading defaults and costs of insolvencies, which are exacerbated
by contemporaneous weaknesses due to correlated portfolios;
\item  via fire sales which depress values of commonly held assets;
\item via uncertainty, fear, and inferences that can result in runs by
depositors and investors, as well as credit freezes in the network itself;
\item via multiple equilibrium values due to coordination in payments.
\item  In some crises, all of these interact at once,  as a bank can be
vulnerable for one reason (e.g., due to a liquidity shortfall and credit freeze, or run, or losses due to depressed asset values due to another bank's fire sale) and can then cause a cascade for another (e.g., default on its liabilities).
\end{itemize}

\item Network structure is essential to understanding contagion and systemic risk:
\begin{itemize}
\item financial networks are ``robust-yet-fragile'':   interconnections help in  diversifying some risks and avoiding individual defaults as banks can insure each other against liquidity
shortfalls, but the interconnectedness makes the network more susceptible to
larger losses in fundamental asset values that then cascade;
\item there is a non-montonicity in how interconnectedness affects contagion:  without any connections there can be no contagion,
with intermediate levels of connections
an institution is heavily exposed to each of its counterparty's default
and
the whole system is at risk, while when institutions have many
counterparties then no individual default is contagious
and contagion becomes less likely;
\item correlation in portfolios can greatly increase the probability of systemic failure and can undo the benefits of having many counterparties;
\item the risk of contagion depends
on the distribution of exposures across banks and  the specific structure of the network (e.g., core-periphery networks act differently than other more balanced networks, being more susceptible to some shocks and less to others);
\item whenever defaults involve some deadweight costs, cycles in the network (of debt, equity, and other contracts)
can generate multiple equilibria and
enable self-fulfilling cascades of defaults;
\item the types of contracts connecting firms (debt, equity, derivatives) matter, and
equity-like contracts can lead to fewer defaults than debt-like contracts all else held equal.
\end{itemize}

\item Financial institutions' investment decisions have externalities that they do not internalize.  They have incentives:
\begin{itemize}
\item to take on more risk than is socially optimal since they see the returns, but do not experience the full costs of defaults, bankruptcies and other
cascading effects;
\item to take on too few counterparties and do too much business with each, inducing excess systemic risk;
\item to correlate risks with their counterparties so that they are solvent when their counterparties pay off on contracts, but are insolvent
when their counterparties default;
\item  and to connect with those others with whom they are most correlated.
\item Reputations and costs of capital can counter some of these incentives, but generally do not fully eliminate these moral hazard problems.
\item In some circumstances, banks have incentives to bail each other out rather than face losses due to another's bankruptcy costs, which can avoid a crisis.
\end{itemize}

\item Different forms of regulation, oversight, and intervention can address different aspects of
systemic risk:
\begin{itemize}

\item Explicit/implicit insurance of contracts and payments can
eliminate fears, runs,  contagion by inference, and credit freezes.

\item Networks are needed to understand and evaluate risks
and to identify optimal ways to intervene, including via bailouts, regulation, and injection of capital:
\begin{itemize}
\item Stress testing should be network-based,  and correlations in portfolios mean that rare events may not be so rare.
\item Understanding  externalities and excess correlation and risk taking requires network information.
\item Potential multiplicity of equilibria requires identifying cycles.
\end{itemize}

\item  When multiple equilibria are present due to cycles, bad equilibria can be eliminated by:
\begin{itemize}
\item `compression' (the canceling out of cycles);
\item the guarantee of payments;
\item restructuring the network
 (e.g., via CCPs), but such `star networks' are vulnerable to failure of the CCPs;
 \item minimal injections of capital, the amounts of which and
points of injection can be fully characterized as a function of cycles in
the network.
\end{itemize}

\item Default cascades can be avoided before they start,  or halted after they start
by combinations of restrictions on investments (e.g. reserve requirements, capital ratios,
restrictions on
correlations of assets with counterparties)  and  bailing out insolvent institutions by making (some of) their payments.
\begin{itemize}
\item  Whether it is better to restrict investments,  insure with ex post bail outs,  or not intervene at all,
depends on the financial centrality of the institutions in question, the network structure, and cost of bailout capital.
\end{itemize}

\item  The relevant measure of `financial centrality' depends on the
specific circumstances and network in question:
\begin{itemize}
\item institutions can be central without being enormous;
\item contracts beyond debt matter and cascades can happen without explicit defaults by all involved
(e.g., a drop in
the equity value of one institution can push another holding that equity into insolvency);
\item correlated investments matter;
\item centrality can be defined directly via the network;
\item different networks interact and a firm might be vulnerable in one network (e.g., via runs and inference from some other failure) and then cause a cascade into another (e.g., then default on its payments), and so full evaluation requires mapping multiple networks.
\end{itemize}
\end{itemize}

\item The financial system is dynamic and reacts to regulation:
\begin{itemize}
\item some investors shift funds to institutions that are less constrained (regulated) and can offer higher returns;
\item new institutions grow outside of regulatory boundaries, and the shadow banking system is large and difficult to observe;
\item investments cross international jurisdictions, making it hard for any regulator to view the network;
\item financial networks are intertwined with international supply chains
that have grown increasingly large and complex.
\end{itemize}

\end{itemize}

\end{document}